\documentclass[11pt,a4paper]{article}
\usepackage[pdfauthor={P. P. Cook, S. Sarkar},
pdftitle={The Two-Parameter Brane sigma-Model and Extended Solutions of M-Theory}]{hyperref}
\usepackage{mathrsfs}
\usepackage[utf8]{inputenc}
\usepackage{amsfonts}
\usepackage{amssymb}
\usepackage{cite}
\usepackage{color,soul}
\usepackage{tabularx}
\usepackage{array}
\usepackage{ragged2e}

\definecolor{lightblue}{rgb}{.90,.95,1}
\definecolor{luridgreen}{rgb}{.4,1,0.7}

\sethlcolor{luridgreen}

\usepackage{amsmath}
\usepackage{graphicx}
\usepackage{chngcntr}
\newcommand{\cb}{\color{black}}
 \newcommand{\cbl}{\color{black}}
\newcommand{\cred}{\color{black}}
\renewcommand{\vec}[1]{{\underline{#1}}}

\makeatother

\begin{document}

\title{The Two-Parameter Brane $\sigma$-Model and Extended Solutions of $M$-Theory}

\author{Paul P. Cook\footnote{email: paul.cook@kcl.ac.uk} and Sarben Sarkar \footnote{email:sarben.sarkar@kcl.ac.uk}}

\begin{titlepage}

\vspace{40pt}
\centering{\LARGE The Two-Parameter Brane $\sigma$-Model and Extended Solutions of $M$-Theory}\\
\vspace{30pt}
\def\thefootnote{\fnsymbol{footnote}}
Paul P. Cook\footnote{\href{mailto:paul.cook@kcl.ac.uk}{email: paul.cook@kcl.ac.uk}}
\\
\vspace{10pt}
{\itshape Department of Mathematics, King's College London \\ 
Strand, London WC2R 2LS, UK}
\vspace{10pt}
\\
and \\
\vspace{10pt} 
Sarben Sarkar\footnote{\href{mailto:sarben.sarkar@kcl.ac.uk}{email: sarben.sarkar@kcl.ac.uk}}
\\
\setcounter{footnote}{0}
\vspace{10pt}
{\itshape Theoretical Particle Physics and Cosmology Group, Department of Physics, King's College London \\ 
Strand, London WC2R 2LS, UK}
\vspace{30pt}
\begin{abstract}
We investigate two-parameter solutions of $\sigma$-models on two dimensional symmetric spaces contained in $E_{11}$. The $\sigma$-models considered here are not proposed as fundamental two-dimensional quantum field theories, but as auxiliary, solution-generating constructions extending the one-parameter brane $\sigma$-model framework. {\cred Embedding such $\sigma$-model solutions in space-time gives classical bosonic solutions of $M^*$- and $M'$-theory whose warp factors are travelling wavefunctions solving a transverse wave equation, rather than the harmonic functions appearing in the standard one-parameter brane solutions.} Weyl reflection allows such solutions to be mapped to M-theory solutions where the wave functions depend explicitly on extra coordinates contained in the fundamental representation of $E_{11}$.
\end{abstract}
\end{titlepage}
\clearpage
\newpage

\section{Introduction}
Shortly after the turn of the century, the Kac–Moody algebra $E_{11}$ was conjectured to encode symmetries of eleven-dimensional $M$-theory \cite{West:2001as}. Around the same time it was shown that the dynamics of eleven-dimensional supergravity near a space-like singularity are encoded in a one-parameter $\sigma$-model invariant under the action of a coset group associated with a Kac-Moody algebra $E_{10}$ \cite{Damour:2002cu,Damour:2002et}. The Lagrangian of the $\sigma$-model is defined in terms of scalar fields on a coset space, and its equations of motion define a null geodesic on the symmetric space from which space-time solutions may be reconstructed. In this sense, the $\sigma$-model should be viewed as a kinematical device encoding algebraic and geometric data, rather than as a fundamental worldsheet theory. In \cite{Damour:2002cu,Damour:2002et} the fields were parameterised by time, a truncation sensible near a space-like singularity but one which obscures the symmetry between spatial and temporal coordinates. The Lorentz symmetry was re-introduced into the $\sigma$-model construction in \cite{Englert:2003py} and used to construct solutions dependent on space-time from $\sigma$-models on symmetric spaces embedded within the Kac-Moody algebra $E_{11}$. This formulation of constructing space-time solutions from sub-algebras of a Kac-Moody algebra has been called the `brane $\sigma$-model' and was used more recently to reconstruct bound state solutions in $M$-theory and string theory \cite{Houart:2009ya,Cook:2011ir} following observations made in \cite{Cook:2009ri}. In all these cases, the solutions of supergravity, superstring theory and $M$-theory were encoded in the path traversed by a massless particle on the coset space. { \cred In the present work we explore the possibility of replacing the point-particle worldline with an auxiliary two-parameter worldsheet on the symmetric space.}

The algebra $E_{11}$ was conjectured to encode the symmetries underlying $M$-theory in \cite{West:2001as}, where this was made manifest via the Borisov–Ogievetsky construction \cite{Borisov:1974bn} (used originally to generate the diffeomorphism algebra of gravity by finding the closure of the conformal group and the Poincar\'{e} group). In the context of $M$-theory the construction involves finding the closure of the conformal group with the bosonic part of the supergravity algebra in eleven dimensions. The appearance of $E_{11}$ in dimensional reduction was anticipated \cite{Julia:1980gr} as the end of the $E_{11-d}$ sequence of `hidden' symmetries that act on the scalar fields of a Kaluza-Klein reduction of bosonic supergravity in eleven dimensions to $d$-dimensions. In other words, $E_{11}$ was expected to arise in the zero-dimensional reduction, but West conjectured that it is already present in an eleven-dimensional extension of supergravity, i.e. $M$-theory \cite{West:2001as}. It was subsequently shown that $E_{11}$ has a very simple relationship to the type IIA and type IIB string theories: the gauge fields of the bosonic parts of these string theories which source the string, the $Dp$-branes as well as the NSNS fields all emerge from the representation theory of $E_{11}$ \cite{Kleinschmidt:2003mf}; the brane solutions of the string theories as well $M$-theory are all straightforwardly encoded in a solution-generating group element \cite{West:2004st} and basic properties relating fundamental dimensionful quantities in each theory have also been derived from the Kac-Moody algebra \cite{West:2001as}. There is now a wealth of literature supporting the $E_{11}$ conjecture in a variety of settings. Furthermore, there is substantial work on the over-extended algebra $E_{10}\subset E_{11}$ whose importance in dynamics in the vicinity of cosmological singularities \cite{Damour:2002cu} which motivated the first investigations of $\sigma$-models in $M$-theory: brane solutions were constructed and $E_{10}$ has been used to construct the fermionic terms expected in supergravity. For a review of $E_{10}$ and cosmological billiards see \cite{Henneaux:2007ej}. In this paper we will work with $E_{11}$. 

It should also be noted that developments in hidden symmetries and dualities in general relativity have stimulated the study of cosets of subgroups of $E_{10}$, $E_{11}$ and other Kac–Moody algebras. In an early work, Buchdahl \cite{BUCHDAHL01011954} noticed a transformation between two static solutions of Einstein's equations which can now be interpreted as T-duality. Subsequently Ehlers \cite{Ehlers:1957zz} uncovered a symmetry which generated solutions of Einstein's equations. A significant breakthrough came with the work of Geroch \cite{Geroch:1970nt,Geroch:1972yt} who extended Ehlers' $SL(2,\mathbb{R})$ symmetry to the (infinite dimensional) affine $SL(2,\mathbb{R})$ Kac-Moody algebra for axi-symmetric stationary solutions. These ideas were developed and used in general relativity for the generation of solutions. Such hidden symmetries were later discovered to be present in supergravity. In particular for $N=8$ supergravity there is a continuous global group $E{_{7(7)}}$ which is a symmetry \cite{Cremmer:1978ds,Cremmer:1979up} - $E{_{7(7)}}$ is a non-compact version of $E_{7}$; it is spontaneously broken in supergravity. These observations led to a broader study of symmetry groups and their generalisations. The symmetry groups above were found for special classes of solutions of Einstein's equation. There has been a quest for larger symmetries associated with the full theory. For generic solutions of Einstein’s equations in four-dimensional spacetime, singularities are expected to occur. For a space-like singularity Belinskii, Khalatnikov and Lifshitz   \cite{belinskiiOscillatoryApproachSingular1970} pointed out that near such a singularity, the spatial metrics at each spatial point are decoupled and they satisfy non-linear second order ordinary differential equations in time. Misner \cite{Misner:1969ae,Misner:1972js} initiated the study of such space-like singularities using Hamiltonian theory and this led to a billiard description \cite{Damour:2002et} in hyperbolic space. Pure gravity billiards have an underlying hidden Kac-Moody algebra as a symmetry. Such symmetries can be studied with the help of geodesic sigma-models. It is through these $\sigma$-models that our approach overlaps with the billiards programme. 

\cb{The two-parameter sigma-model used below is not proposed as a
fundamental worldsheet quantum field theory.  Its role is analogous to
that of the one-parameter brane sigma-model: it is an auxiliary
solution-generating device which probes how the algebraic data of
symmetric spaces embedded in $E_{11}$ are lifted to spacetime fields.
{\cred The physical information extracted from the construction concerns the
classical solution space of the $E_{11}$/$M$-theory framework and its
$M^\ast$ and $M'$ signature frames, rather than quantum observables
of a fundamental worldsheet theory.}

The symmetric spaces used in brane $\sigma$-models are constructed using the Kac-Moody algebras $E_{10}$ and $E_{11}$. These algebras have long been argued to encode hidden symmetries of supergravity relevant to $M$-theory \cite{Julia:1980gr,West:2001as} and the coset construction provides a dictionary relating the path of a massless particle on a coset to brane solutions in space-time. The $\sigma$-model action on the coset for a massless particle is a simpler and, arguably, more fundamental setting to investigate $M$-theory and the high-energy description of space-time. It is therefore of interest to consider simple extensions of massless particle motion on cosets of subgroups of $E_{10}$ and $E_{11}$. {\cred Here, instead of massless particle motion, we consider an auxiliary two-parameter worldsheet construction on cosets of subgroups of $E_{11}$.} The central problem is how to embed two symmetric-space parameters in space-time, i.e. how to associate coordinates on the symmetric space with space-time coordinates.

The approach to the one-parameter brane-$\sigma$-model established in the literature \cite{Damour:2002cu,Damour:2002et,Englert:2003py,Houart:2009ya,Cook:2011ir} successfully constructs solutions of supergravity and string theory\footnote{We will focus on the supergravity solutions in the following, but similar comments are relevant to the string theory solutions too.} dependent on one space-time coordinate and involves several steps:
\begin{enumerate} 
\item identifying a symmetric space $\frac{G}{{\cal K}(G)}$ which can be embedded within $\frac{E_{11}}{{\cal K}(E_{11})}$ or $\frac{E_{10}}{{\cal K}(E_{10})}$,
\item solving the equations of motion of the sigma-model Lagrangian with the symmetries of $\frac{G}{{\cal K}(G)}$ to find scalar fields\footnote{The number of fields indicated by the range of $i$ and $I$ is determined by the Borel algebra of $G$.} $\phi_i(\xi)$ and $C_I(\xi)$ dependent on a single parameter $\xi$, 
\item identifying $\xi$ with a space-time coordinate, and 
\item upgrading the scalar fields of the sigma-model to tensor fields of supergravity (i.e. the eleven-dimensional vielbein and the components of the four-form field strength) by embedding the symmetric space $\frac{G}{{\cal K}(G)}$ in $\frac{E_{11}}{{\cal K}(E_{11})}$ or $\frac{E_{10}}{{\cal K}(E_{10})}$.
\end{enumerate}
The validity of step three lies in its ability to reconstruct known supergravity solutions, which have been established in the literature: for example, all $\tfrac{1}{2}$-BPS branes of supergravity and string theory \cite{Englert:2003py} have been constructed in this way from $E_{11}$ as have bound states of branes in both supergravity and string theory \cite{Cook:2009ri,Houart:2009ya,Cook:2011ir}. In all these examples, the identification of $\xi$ with a space-time coordinate is \emph{a priori} unexpected but is confirmed once, on doing so, it leads to a known supergravity solution.

The symmetric spaces related to single brane solutions are two-dimensional manifolds and our aim is to investigate two-parameter solutions on these manifolds and consider their embedding in space-time. The $\sigma$-model we will investigate predominantly is defined on $\frac{SL(2,\mathbb{R})}{SO(1,1)}$ a pseudo-Riemannian  symmetric space. For simplicity our investigations will be restricted to a small neighbourhood of the identity on the symmetric space, allowing the approximation of using the flat space Minkowski metric on the symmetric space. One consequence of developing two-parameter solutions is that the embedding space (usually taken to be the transverse space to the brane world-volume) will contain both timelike and spacelike coordinates. The resulting two-parameter metrics and gauge-fields will be solutions of $M^*$ and $M'$ theories \cite{Hull:1998ym}. The $E_{11}$ conjecture leads to an enhancement of $M$-theory which contains both $M^*$ and $M'$-theories \cite{Keurentjes:2005jw}: sub-algebras of $E_{11}$ related to the three theories are mapped into each other under $E_{11}$ Weyl reflections. Since $\sigma$-model solutions are preserved under Weyl reflections, this raises the question: what do the two-parameter solutions of $M^*$- and $M'$-theories (which are characterised by wave-functions of the transverse coordinates) map to in $M$-theory (where solutions are characterised by harmonic functions of the transverse space coordinates)? We will argue that $M^*$ and $M'$-theory solutions defined in terms of the coordinates $x^\mu$ are mapped to $M$-theory solutions which depend on the extra $E_{11}$ coordinates $y_{\mu_1\mu_2}$, $z_{\mu_1\mu_2\mu_3\mu_4\mu_5}$, $w_{\mu_1\mu_2\mu_3\mu_4\mu_5\mu_6\mu_7|\nu}$, $w_{\mu_1\mu_2\mu_3\mu_4\mu_5\mu_6\mu_7\mu_8}$ and so on which are contained in the $l_1$ representation of $E_{11}$ \cite{Kleinschmidt:2003jf}. These are the same extra coordinates that are central to the internal symmetries of double field theory \cite{Siegel:1993th,Hull:2009mi} and exceptional field theory \cite{Hohm:2013pua} which were first introduced in \cite{Duff:1989tf,Duff:1990hn}. {\cred Recent developments in $E_{11}$, E-theory and exceptional field theory have further clarified the role of extended spacetime, $l_1$ coordinates, section-condition-like constraints and brane-related dependence on coordinates beyond the usual spacetime coordinates \cite{boulangerUnfolding$E_11$2025,westLocalSymmetryExtended2025,samtlebenExceptionalFieldTheories2025,ostenUniversalExceptionalStructure2024}.} The extra coordinates will require an extension of the brane $\sigma$-model which we will discuss in our concluding remarks.

\cb{The novelty of the present construction is not the use of a symmetric
space by itself, since symmetric-space sigma-models have long appeared
in the study of hidden symmetries, cosmological billiards and brane
sigma-models.  The new point is that the usual one-parameter null
geodesic is replaced by a two-parameter worldsheet on the same
symmetric space.  This replacement changes the spacetime functions
obtained after embedding: the harmonic functions of the standard
brane solutions are replaced by travelling wavefunctions.  The
embedding of these solutions forces a transverse sector with
$SO(1,1)$ signature and hence naturally leads to $M^\ast$- and
$M'$-theory.  Weyl reflection then maps these solutions to the
ordinary M-theory duality frame, where their wavefunction dependence
is carried by extended $E_{11}$ coordinates.  {\cred This gives a concrete solution-level argument for the relevance, within the $E_{11}$ framework, of the extended coordinate system.}
}\cbl

This paper is organised as follows: In section \ref{1dbranesolutions} we review the construction of $\frac{1}{2}$-BPS brane solutions of supergravity as null geodesic motions of a particle on cosets of $SL(2,\mathbb{R})$ embedded in $E_{11}$. Our aim in this section is to familiarise the reader with the one-parameter brane $\sigma$-model before presenting the two-parameter model in section \ref{2dsigmamodel}. We will then present simple solutions to the two-parameter model, some of which are only solutions in two-dimensions, and other more general solutions which are described by wavefunctions. In section \ref{Mtheories} we will embed the two-parameter solutions into eleven-dimensional space-times with multiple time coordinates and show some examples of solutions to bosonic $M^*$-theory and $M'$-theory described in terms of wavefunctions. In particular we will present some novel cosmological solutions in a background of two time coordinates which evolve so as to suppress the second temporal coordinate. We will apply Weyl reflections to map these solutions into $M$-theory and see that they depend on extra coordinates. We conclude in section \ref{conclusions}.

\section{Brane solutions and the brane $\sigma$-model}\label{1dbranesolutions}

In this section we will review the one-parameter brane $\sigma$-models and show how given an involution $\Omega$ defined on the algebra of $E_{11}$ we may define the ``group" ${\cal K}(E_{11})$ whose algebra is fixed under $\Omega$. We will then review how any real root of $E_{11}$ may be associated with a sigma-model on the symmetric spaces $\frac{SL(2,\mathbb{R})}{SO(1,1)}$ or $\frac{SL(2,\mathbb{R})}{SO(2)}$. In the former case the solutions of the equations of motion encode electric brane solutions of $M$-theory, or one of its counterparts $M^*$ or $M'$-theory \cite{Hull:1998ym}, while there are no real solutions to the equations of motion in the latter case. Our principal example will be the solution of the $\frac{SL(2,\mathbb{R})}{SO(1,1)}$ model associated with the exceptional root $\vec{\alpha}_{11}$ of $E_{11}$, which encodes the $M2$-brane of $M$-theory. We commence by defining the involution $\Omega$ which leaves the algebra of ${\cal{K}}(E_{11})$ invariant before constructing the $\sigma$-models on $\frac{SL(2,\mathbb{R})}{SO(1,1)}$ and $\frac{SL(2,\mathbb{R})}{SO(2)}$. The involution $\Omega$ encodes the signature of the background space-time. The equations of motion on each symmetric space will be presented, and it will be observed that the $\sigma$-model on the first space has real solutions which will be embedded into space-time while the $\sigma$-model on the second space does not possess real solutions.

\subsection{The signature of space-time.}\label{temporalinvolution}
The algebra $E_{11}$ is infinite dimensional, and is associated with the extension of eleven-dimensional supergravity by singling out an $SL(11,\mathbb{R})$ sub-algebra. This sub-algebra is formed of the nodes $1,2\ldots 10$ labelled on the Dynkin diagram of $E_{11}$ as depicted in figure \ref{e11} and sometimes called the gravity line.
\begin{figure}[h]
\centering \includegraphics[scale =.7]{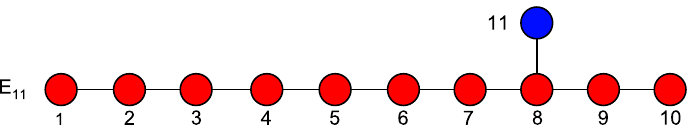}
\caption{\small The Dynkin diagram of $E_{11}$, the line of red nodes indicate the Dynkin diagram of $SL(11,\mathbb{R})$ associated with the eleven dimensional space-time.} \label{e11}
\end{figure}
The decomposition of the algebra $E_{11}$ with respect to the gravity line gives an infinite set of tensor representations of $SL(11,\mathbb{R})$. Any root $\vec{\beta}$ within the root space of $E_{11}$ may be written as a sum of simple, positive roots: $\vec{\beta}=\sum_{i=1}^{11}m_i\vec{\alpha}_i$ and the decomposition into representations of $SL(11,\mathbb{R})$ may be understood as splitting the root $\vec\beta$ into a weight in the weight space of $SL(11,\mathbb{R})$, denoted $\vec\Lambda$, and a part $\vec{x}$ orthogonal to the weights of $SL(11,\mathbb{R})$, i.e. $\vec{\beta}=\vec\Lambda+m_{11}\vec{x}$. The coefficient $m_{11}$ is called the level and labels sets of representations of $SL(11,\mathbb{R})$ defined by a lowest (equivalent to the highest weight labelling) weight $\vec\Lambda$. For example at level $m_{11}=0$, we find the roots of $SL(11,\mathbb{R})$, at level one ($m_{11}=1$) we find an antisymmetric three-tensor representation $R^{a_1a_2a_3}$, at level two we have an antisymmetric six-tensor $R^{a_1\ldots a_6}$, at level three we find a mixed symmetry tensor $R^{a_1\ldots a_8|b}$ and so on.

The signature of space-time is derived from the quadratic form invariant under $SO(1,10)$. The algebra of this group is invariant under the (temporal) involution $\Omega$ which acts on the positive generators $E_i$ of $\mathfrak{sl}(11,\mathbb{R})$ as 
\begin{equation}
\Omega(E_i)=-\epsilon_i(F_i), \label{involution}
\end{equation}
where $F_i$ are the negative generators and $\epsilon_i=\pm 1$ where $i\in\{1,2,\ldots 10\}$. The sub-algebra of $\mathfrak{sl}(11,\mathbb{R})$ invariant under $\Omega$ is $\mathfrak{so}(p,q)$ where $p,q\in\mathbb{Z}_0^+$ such that $p+q=11$\footnote{The values of $p$ and $q$ are determined from the involution parameters $\epsilon_i$ or equivalently the signature function $\vec{f}$ which will be defined later in this paper.}, having generators $Q_i\equiv E_i-\epsilon_i F_i$ (where there is no summation over the repeated indices). The remainder of the algebra (the anti-fixed set of generators under $\Omega$) consists of $P_i\equiv E_i+\epsilon_i F_i$ and the generators of the Cartan sub-algebra $H_i$. For example if $\epsilon_1=\epsilon_2=\ldots \epsilon_{10}=1$ then the sub-algebra fixed under $\Omega$ is spanned by the generators $Q_i\equiv E_i- F_i$, or in other words $\mathfrak{so}(11)$. Alternatively if $\epsilon_1=-1$, while $\epsilon_2=\ldots \epsilon_{10}=1$ the fixed sub-algebra is $\mathfrak{so}(1,10)$. By appropriate choice of $\epsilon_i$ the sub-algebra $\mathfrak{so}(t,11-t)$ may be constructed. Extending the action of $\Omega$ to the generator $R^{9\,10\,11}$ associated with the exceptional node of the $E_{11}$ Dynkin diagram (the blue eleventh node shown in figure \ref{e11}) as $\Omega(R^{9\,10\,11})=\epsilon_{11}R_{9\,10\,11}$, one may define ${\cal K}(E_{11})$ as the exponentiation of the sub-algebra invariant under the involution $\Omega$. The role of $\epsilon_{11}$ is to control the sign in front of the kinetic term for the three-form gauge field in the supergravity action \cite{Keurentjes:2004bv}. In summary $\Omega$ is defined by a choice of eleven numbers $\epsilon_i$ and encodes the space-time signature and the sign of the kinetic term in the action. 

It was shown in \cite{Keurentjes:2004bv} that the signature of space-time defined by the choice of $\Omega$ is not invariant under the Weyl reflections of $E_{11}$. This may be understood as follows: the signature of space-time depends on the action of $\Omega$ on the generators of $\mathfrak{sl}(11,\mathbb{R})$ given by the gravity line (the red nodes numbered from one to ten in figure \ref{e11}), but under a Weyl reflection the generators associated with the gravity line may be mapped to another $\mathfrak{sl}(11,\mathbb{R})$ sub-algebra within $E_{11}$ and vice-versa, so that the gravity line is formed of a different set of $\mathfrak{sl}(11,\mathbb{R})$ generators before and after a Weyl reflection. Consequently the sub-algebra of the gravity line algebra which is invariant under $\Omega$ may change under a Weyl reflection. For example, commencing with a choice of $\Omega$ which leaves $\mathfrak{so}(1,10)$ fixed within the gravity line algebra\footnote{To give the space-time signature relevant to supergravity in eleven dimensions.} following an $E_{11}$ Weyl reflection the gravity line sub-algebra fixed by $\Omega$ may be one of $\mathfrak{so}(1,10)$, $\mathfrak{so}(2,9)$, $\mathfrak{so}(5,6)$, $\mathfrak{so}(6,5)$, $\mathfrak{so}(9,2)$ or $\mathfrak{so}(10,1)$ \cite{Keurentjes:2004bv}. This potential change of signature highlights the unnatural manner in which $E_{11}$ is associated with an eleven-dimensional space-time by an essentially arbitrary choice of $\mathfrak{sl}(11,\mathbb{R})$ sub-algebra. It is more natural to consider $E_{11}$ as the symmetries of a theory on an infinite dimensional space-time whose coordinates parametrise the vector representation of $E_{11}$ and whose isometry group is ${\cal K}(E_{11})$. In this paper we will adopt this viewpoint and will aim to identify coordinates on $\frac{E_{11}}{{\cal K}(E_{11})}$ directly with coordinates in space-time. A Weyl reflection in such a setting would leave the isometry group unaltered and would map active fields in one sector to another, preserving solutions. With the restriction to an eleven dimensional space-time the effect of a Weyl reflection is to change the signature of space-time while preserving solutions derived in different sectors of the theory. 

While $\epsilon_i$ for $i=1,\ldots 10$ define the signature of space-time, $\epsilon_{11}$ is associated with the sign of the four-form kinetic term \cite{Keurentjes:2004bv}. We will describe the dependence of the sign of this term on $\epsilon_{11}$ via the signature function, $\vec f$, introduced in \cite{Keurentjes:2004bv}. The signature function is a function on the weight space of $E_{11}$ defined from the involution $\Omega$ by 
\begin{equation}
\epsilon_k=e^{i\pi<\vec\alpha_k,\vec{f}>}
\end{equation}
where upon writing $\vec{f}\equiv \sum p_i\vec\lambda_i$, $\vec\lambda_i$ are the fundamental weights of $E_{11}$, we have $\epsilon_k=e^{i\pi p_k}$ and as $\epsilon_k\in \{-1,1\}$ then $p_k\in\{0\mod 2, 1\mod 2\}$. The Weyl reflections in real roots $\vec\beta$ of $E_{11}$ denoted $W_{\vec\beta}$ and defined by
\begin{equation}
W_{\vec\beta}(\vec{\alpha})=\vec{\alpha}-2\frac{<\vec{\alpha},\vec{\beta}>}{<\vec{\beta},\vec{\beta}>}{\vec{\beta}}
\end{equation}
map the signature function from $\vec f$ to $\hat{f}\equiv W_{\vec{\beta}}(\vec f)$. 

Throughout this paper we will work with a truncation of $E_{11}$ to generators $H_{\alpha_{11}}$, $E_{\alpha_{11}}$ and $F_{\alpha_{11}}$. This sub-algebra is related to the gravity and the three-form sectors of supergravity. It will be helpful to establish the effect of a Weyl reflection on the signs of the terms in this part of the supergravity action. The conventions we adopt are as follows
\begin{equation}
S=\int R\star 1 - (-1)^{(f_0+<\vec{f},\vec\gamma>)} \frac{1}{2} F_{\alpha_{11}}\wedge \star F_{\alpha_{11}} +\ldots \label{genericlowlevelaction}
\end{equation}
where $f_0\equiv 1$ if $x^1$ is a temporal coordinate and $f_0\equiv 0$ if $x^1$ is a spatial coordinate and $\vec\gamma\equiv \vec{\alpha}_2+\vec{\alpha}_4+\vec{\alpha}_6+\vec{\alpha}_8+\vec{\alpha}_{10}+\vec{\alpha}_{11}$; $F_{\alpha_{11}}$ is the field strength sourced by a membrane oriented along $\{x^9,x^{10},x^{11}\}$ and the ellipsis denotes other terms. We have focussed on the kinetic term relevant to the membrane oriented along $\{x^9,x^{10},x^{11}\}$. The sign of the term is derived from the number of timelike coordinates among the directions $\{x^9,x^{10},x^{11}\}$: if the number is odd the term should be negative, if even the term should be positive. Supposing that $x^8$ is spatial then the crucial number $<\vec{f},3\vec\alpha_8+2\vec\alpha_9+\vec\alpha_{10}>$, which counts the number of time directions modulo two along the brane world-volume, may be simplified, modulo two, to $<\vec{f},\vec\alpha_8+\vec\alpha_{10}>$. A similar construction can be used to account for $x^8$ being either spatial or temporal, modulo two we have $-(-1)^{<\vec{f},\vec{\alpha}_2+\vec{\alpha}_4+\vec{\alpha}_6+\vec{\alpha}_8+\vec{\alpha}_{10}>}$. Together with an additional minus sign given by $(-1)^{<\vec{f},\vec\alpha_{11}>}$, we have $-(-1)^{<\vec{f},\vec\gamma>} \frac{1}{2} F\wedge \star F$. The signature function keeps track of all the times a successive coordinate switches from being temporal to spatial a final parameter $f_0$ is required to specify whether the first coordinate is spatial or temporal which affects the overall sign of the kinetic term to give, in all, $-(-1)^{f_0+<\vec{f},\vec\gamma>} \frac{1}{2} F\wedge \star F$. \footnote{\label{footnote_on_signs}Later in this paper we will consider Taub-NUT solutions rather than membrane solutions. The Taub-NUT solution is more involved than the membrane solution, it appears in $E_{11}$ via a mixed-symmetry gauge field $A_{\mu_1\ldots \mu_8|\nu}$ which is dual to the vielbein in eleven dimensions. In this presentation the Taub-NUT sources a field strength $F_{TN}$ with kinetic term is $- (-1)^{(f_0+<\vec{f},\vec\alpha_2+\vec\alpha_3+\vec\alpha_5+\vec\alpha_7+\vec\alpha_{9}+\vec\alpha_{11})>} \frac{1}{2} F_{TN}\wedge \star F_{TN}$.} For example the signature function $\vec{f}=\lambda_{10}+\lambda_{11}$ where $x^1$ is space-like (so $f_0=0$) corresponds to the signature $(1,10)$ with $x^{11}$ being the temporal coordinate and $S=\int R\star 1 - \frac{1}{2} F_{\alpha_{11}}\wedge \star F_{\alpha_{11}} +\ldots $. Under a Weyl reflection the signature function is mapped from $\vec f$ to $\hat{f}\equiv W_{\vec{\beta}}(\vec f)$. This has the consequence of not only changing the signature of space-time but also the sign of the kinetic term in the action in equation (\ref{genericlowlevelaction}). The involution $\Omega$ and hence the algebra of ${\cal K}(E_{11})$, or equivalently the signature function $\vec{f}$, is the first ingredient which must be specified before the brane $\sigma$-model may be constructed.

\subsection{The one-parameter brane $\sigma$-model}
For $G$, a semisimple Lie group embedded in $E_{11}$, $\sigma$-models constructed on the symmetric space $\frac{G}{{\cal K}(G)}$ have solutions which encode half-BPS brane solutions \cite{Englert:2003py} and bound states of these brane solutions \cite{Cook:2009ri,Houart:2009ya,Cook:2011ir}. We restrict our attention in this paper to single brane solutions which correspond to identifying $G=SL(2,\mathbb{R})$ embedded in $E_{11}$. The truncation of the algebra $E_{11}$ to $\mathfrak{sl}(2,\mathbb{R})$ gives an associated truncation of the algebra of ${\cal K}(E_{11})$ to ${\cal K}(SL(2,\mathbb{R}))$ which is defined using the involution $\Omega$ on $E_{11}$. Taking $G=SL(2,\mathbb{R})$ whose single positive root $\vec{\beta}$ is a real root in the root lattice of $E_{11}$, ${\cal K}(G)$ is either $SO(2)$ or $SO(1,1)$. Given a signature function $\vec f$ then if $<\vec f,\vec\beta>=1\mod{2}$ then ${\cal K}(G)=SO(1,1)$, while if $<\vec f,\vec\beta>=0\mod{2}$ then ${\cal K}(G)=SO(2)$ \cite{Keurentjes:2004bv}. We will observe in the construction of the $\sigma$-models how the sign choices of the kinetic term of the action in eleven dimensions above are related to the signs appearing in the one-dimensional action of the $\sigma$-models.

\subsubsection{The Lagrangian density}
The one-parameter $\sigma$-model has an action which is invariant under the symmetries of the coset $\frac{G}{{\cal K}(G)}$ and is defined by 
\begin{equation}
S=\int d\xi \, {\cal L}
\end{equation}
where $\xi$ is a single coordinate on the coset manifold on which the Lagrangian density ${\cal L}$ depends. The Lagrangian density is defined in terms of an inner product by
\begin{equation}
{\cal L}=\eta^{-1}(P_\xi |P_\xi)
\end{equation}
where $P_\xi$ is derived by decomposing the Maurer-Cartan form $\nu\equiv dgg^{-1}$ for $g\in G$ as follows
\begin{equation}
\nu_\xi =(\partial_\xi g)g^{-1}=P_\xi + Q_\xi \in \mathfrak{g},
\end{equation}
the inner product is the Cartan-Killing form, the generators denoted $Q_\xi$ are in the algebra of ${\cal K}(G)$, $P_\xi$ are the complementary generators in the Lie algebra of $G$ and the field $\eta$ is included to guarantee the reparameterisation invariance of the action $S$.

The Lagrangian density ${\cal L}$ is invariant under the symmetry transformations of the coset. Specifically the global transformation
\begin{equation}
g\rightarrow gg_0
\end{equation}
where $g_0$ is independent of coordinates on the coset manifold leaves the Maurer-Cartan form unchanged:
\begin{equation}
\nu_\xi\rightarrow \partial_\xi(gg_0)(gg_0)^{-1}=\nu_\xi.
\end{equation}
While the transformation under the local sub-group element $k(\xi)\in {\cal K}(G)$ given by
\begin{equation}
g\rightarrow kg
\end{equation}
leaves the Lagrangian density ${\cal L}$ unchanged as
\begin{equation}
\nu_\xi\rightarrow \partial_\xi(kg)(kg)^{-1}=\partial_\xi(k)k^{-1}+k\nu_\xi k^{-1}
\end{equation}
hence
\begin{equation}
P_\xi\rightarrow k P_\xi k^{-1}\qquad \mbox{and} \qquad Q_\xi\rightarrow k Q_\xi k^{-1}+\partial_\xi(k)k^{-1},
\end{equation}
which leaves ${\cal L}$ unchanged. We now construct the Lagrangians for $G=SL(2,\mathbb{R})$.

\subsection{The $\frac{SL(2,\mathbb{R})}{SO(1,1)}$ brane $\sigma$-model.}\label{oneparameterequations}
Let $g\in \frac{SL(2,\mathbb{R})}{SO(1,1)}$ be the coset representative written, in the Borel gauge (upper triangular gauge), in terms of $H$, the Cartan sub-algebra element of the algebra $sl(2,\mathbb{R})$ and $E$, the positive generator of $sl(2,\mathbb{R})$, as follows:
\begin{equation}
g=\exp(\phi H)\exp(CE)\label{borelgauge}
\end{equation}
where $\phi\equiv\phi(\xi)$ and $C\equiv C(\xi)$. $H$ and $E$ are simply represented by two-by-two matrices:
 
\begin{equation}
H=\left(\begin{array}{cc}
1 & 0 \\
0 & -1
\end{array}\right) \qquad \mbox{and} \qquad
E=\left(\begin{array}{cc}
0 & 1 \\
0 & 0
\end{array}\right).
\end{equation}
Hence
\begin{equation}
g=\left(\begin{array}{cc}
e^\phi & e^\phi C \\
0 & e^{-\phi}
\end{array}\right)
\end{equation}
and therefore
\begin{equation}
\nu_\xi=\partial_\xi \phi H + \partial_\xi C \exp(2\phi) E.
\end{equation}
The $so(1,1)$ sub-algebra is the part of $sl(2,\mathbb{R})$ which is invariant under the involution $\Omega$ defined by $\Omega(H)=-H$ and $\Omega(E)=F$ where $F\equiv E^T$ is the negative (lower triangular) generator of $sl(2,R)$. The sub-algebra of $so(1,1)$ contains a single generator $q=\frac{1}{2}(E+F)$, the remainder of the algebra of $sl(2,R)$ is spanned by $H$ and $p\equiv\frac{1}{2}(E-F)$. We have normalised $q$ and $p$ so that $E=q+p$. We have
\begin{equation}
\nu_\xi=\partial_\xi \phi H + \partial_\xi C \exp(2\phi) (p+q)\equiv P_\xi+Q_\xi
\end{equation}
and 
\begin{equation}
P_\xi=\left(\begin{array}{cc} \partial_\xi \phi & \frac{1}{2}e^{(2\phi)} \partial_\xi C\\
-\frac{1}{2}e^{(2\phi)} \partial_\xi C & -\partial_\xi \phi
\end{array}\right).
\end{equation}
This gives us the following Lagrangian density
\begin{equation}
{\cal L}=\eta^{-1}(2(\partial\phi)^2 -\frac{1}{2}e^{4\phi}(\partial C)^2)
\end{equation}
whose equations of motion are
\begin{equation}
\partial^2 \phi +\frac{1}{2}(\partial C)^2 e^{4\phi}=0\, , \quad \partial(\partial C e^{4\phi})=0\quad \mbox{and}\quad
(\partial\phi)^2 -\frac{1}{4}e^{4\phi}(\partial C)^2=0. \label{nullgeodesic}
\end{equation}
These are the equations of motion for $\phi$, $C$ and $\eta$ respectively. 

The final equation of motion above may be written as $(P_\xi|P_\xi)=0$ and hence the path of $P_\xi$ described on the group manifold will be a null geodesic on the coset. The solution will therefore be described by a single parameter, which we have labelled $\xi$. Let us see how the solution emerges.

As $\partial(\partial C e^{4\phi})=0$, then $(\partial C)e^{4\phi}=A$ where $A$ is a constant. Hence $(\partial C)=Ae^{-4\phi}$ and substitution of this identity into the equation of motion for $\eta$ gives
\begin{equation}
(\partial\phi)^2-\frac{1}{4}A^2e^{-4\phi}=0.
\end{equation}
Hence $\partial\phi=\pm \frac{1}{2}Ae^{-2\phi}$. Solutions of this equation are found by integration. The solutions are $\phi=\frac{1}{2}\ln(N)$ where $N=\pm a\xi+b$ is a harmonic function in one of the coset coordinates and $a$ and $b$ are real constants. Substituting $\phi=\frac{1}{2}\ln(N)$ into the equation of motion for $\phi$ and $C$ gives 
\begin{equation}
-\frac{(\partial N)^2}{2N^2}+\frac{1}{2}(\partial C)^2N^2=0 \qquad \mbox{and}\qquad (\partial C)N^2=A.
\end{equation}
Hence $C=-N^{-1}+B$ solves both equations where $B$ is a constant and $A=\partial N=a$. 

\subsubsection{Example: The M2 branes}
Let the algebra $\mathfrak{sl}(2,\mathbb{R})$ used in the coset construction have the following embedding in $E_{11}$:
\begin{align}
E&=E_{\alpha_{11}}\equiv R^{9\,10\,11}, \quad F=E_{-\alpha_{11}}\equiv R_{9\,10\,11} \quad \mbox{and} \label{M2embedding1}\\
H&=H_{11}=-\frac{1}{3}({K^1}_1+\ldots +{K^8}_8)+\frac{2}{3}({K^9}_9+{K^{10}}_{10}+{K^{11}}_{11}).\label{M2embedding2}
\end{align}
A dictionary is used to construct the bosonic part of the brane solution in supergravity. The dictionary is defined in a natural way: active components of the four-form field strength written in flat space are related to the field $C$ in the coset construction by $F_{\xi9\,10\,11}\equiv e^{2\phi}\partial_\xi C$ (the index structure on the field strength is inherited from the index structure of the $E_{11}$ generator, i.e. $CE=C_{9\,10\,11}R^{9\,10\,11}$) and the diagonal components of the elfbein are related to $\phi$ by ${e_\mu}^m\equiv \exp(-\phi {h_m}^m)$ where repeated indices are not summed over and ${h_m}^m$ is defined by $H_{\alpha_{11}}=\sum_m {h_m}^m {K^m}_m$. The coset parameter $\xi$ is embedded in the space-time such that the four-form structure of $F_{\xi9\,10\,11}$ is respected, i.e. $\xi$ may be chosen to be any of the eight coordinates ${x^1,\ldots x^8}$. There remains the choice of the time coordinate which may be on the brane world-volume or transverse to it.\footnote{This is independent of the action of the temporal involution on $E_{\alpha}$, which is used to define the coset: this is always given by $\Omega(E_{\alpha})=E_{-\alpha}$ for the coset $\frac{SL(2,\mathbb{R})}{SO(1,1)}$.}

\subsubsection{The electric brane}\label{M2brane} For this example, without loss of generality, we will choose\footnote{This corresponds to picking the signature function to be $\vec f= -\vec{\lambda_8}+\vec{\lambda_9}+\vec{\lambda_{11}}$ with $f_0=0$.} $t=x^9$ and $\xi=x^1$. This gives the metric:
\begin{equation}
ds^2=N^{\frac{1}{3}}(d\Sigma_8^2)+N^{-\frac{2}{3}}(-dt^2+(dy^1)^2+(dy^2)^2)
\end{equation}
where $d\Sigma_8^2=\sum_{i=1}^8 (dx^i)^2$ and $N=Ax^1 +B$. The non-trivial four-form field strength components are $F_{\hat{1}\,9\,10\,11}\equiv AN^{-1}$ and its anti-symmetrisations. Where we use a hat to differentiate between curved and flat space indices, the hat denoting a curved space index. The field strength is embedded in the curved space-time using the elfbein, so that \begin{equation}
F_{\hat{1}\,\hat{9}\,\hat{10}\,\hat{11}}= {e_{\hat{9}}}^9{e_{\hat{10}}}^{10}{e_{\hat{11}}}^{11} F_{\hat{1}\,9\,10\,11}=-\partial_{\hat{1}}N^{-1}. 	
\end{equation}
For the resulting space-time to be asymptotically Minkowski space corresponds to the limit $\lim_{x^1\rightarrow 0}( g_{\mu\nu})=\eta_{\mu\nu}$, i.e. that $B=1$. Integrating $F_{\mu_1\mu_2\mu_3\mu_4}$ over a spatial seven-sphere leads us to interpret $A$ as the electric charge due to the presence of a membrane, hence we will write $A\equiv {\cal Q}$.

The resulting solution differs from the supergravity membrane solution \cite{Duff:1990xz} as the harmonic function $N=1+{\cal Q}x^{\hat{1}}$ depends on one coordinate in the transverse space. The supergravity membrane solution has an $SO(8)$ isometry in its transverse space, which is not respected by the dependence of $N$ on only $x^{\hat{1}}$. The choice of embedding $\xi$ in space-time as $x^{\hat{1}}$ was arbitrary, any of the eight transverse $x^{\hat{i}}$ ($i\in\{1,2,\ldots 8\}$) would have been as effective. To lift the one-parameter solution to eleven dimensions one has to ensure that $N$ respects the $SO(8)$ symmetry, so that $\xi=r$ where $r^2=(x^{\hat{1}})^2+(x^{\hat{2}})^2+\ldots +(x^{\hat{8}})^2$ and $N$ remains harmonic: $\partial^{\hat{i}}\partial_{\hat{i}} N=0$. This leads to $N=1+\frac{{\cal Q}}{r^6}$: in this manner the one-parameter $M$-theory solution is ``unsmeared" to an eleven dimensional solution.

\subsubsection{The magnetic brane: a no-go condition} We might expect that one can choose a signature function such that $t$ is transverse to the brane. However such a choice is prohibited by ensuring that Poincar\'e duality is consistent for the theory \cite{Keurentjes:2005jw}. A simple condition found in \cite{Keurentjes:2005jw} for a solution which respects the Poincar\'e duality is that 
\begin{equation}
i_0(\vec{f})\equiv<\vec{\alpha_1}+\vec{\alpha_3}+\vec{\alpha_5}+\vec{\alpha_7}+\vec{\alpha_{11}},\vec{f}>=1\mod{2}. \label{i0}
\end{equation}
For any choice of signature function such that the space-time signature is $(1,10)$, the kinetic term has the usual sign $-F^2$ in the action, and the temporal coordinate, $t$, is transverse to the brane world volume, i.e. $t\in\{x^1,\ldots x^8\}$, it may be verified that (\ref{i0}) is not satisfied. 

\subsection{The $\frac{SL(2,\mathbb{R})}{SO(2)}$ brane $\sigma$-model}
Let $g\in \frac{SL(2,\mathbb{R})}{SO(2)}$ be the coset representative written in the Borel gauge (upper triangular gauge) as defined in equation (\ref{borelgauge}). The procedure to construct the $\sigma$-model Lagrangian density is the same as in the preceding section, the only change is that the sub-algebra $so(2)$ is generated by the algebra element $q'=E-F$ while the remaining part of the algebra is spanned by the Cartan element $H$ and $p'=E+F$ (i.e. $q'=p$ and $p'=q$ compared to the previous section). Computation of the Maurer-Cartan form allows $P_\mu$ to be read off as 
\begin{equation}
P_\xi=\left(\begin{array}{cc} \partial_\xi \phi & \frac{1}{2}e^{(2\phi)} \partial_\xi C\\
\frac{1}{2}e^{(2\phi)} \partial_\xi C & \partial_\xi \phi
\end{array}\right).
\end{equation}
This gives a change in sign of the kinetic term for $C$ in the $\sigma$-model Lagrangian density. We now have
\begin{equation}
{\cal L}=\eta^{-1}(2(\partial\phi)^2 +\frac{1}{2}e^{4\phi}(\partial C)^2)
\end{equation}
and the equations of motion are
\begin{equation}
\partial^2 \phi -\frac{1}{2}(\partial C)^2 e^{4\phi}=0, \quad \partial(\partial C e^{4\phi})=0 \quad \mbox{and}\quad
(\partial\phi)^2 +\frac{1}{2}e^{4\phi}(\partial C)^2=0.
\end{equation}
These equations are not solved by the ansatz $\phi=\frac{1}{2}\ln(N)$ where $C$ is a real field. From the second equation, we have $(\partial C) e^{4\phi}=A$, a constant, so the last equation becomes
\begin{equation}
(\partial\phi)^2+\frac{1}{2}A^2 e^{-4\phi}=0
\end{equation}
which has no non-trivial real solution for $\partial \phi$..

\section{The Two-Parameter Brane $\sigma$-model}\label{2dsigmamodel}
We will set out the two-parameter brane $\sigma$-models and equations of motion for the symmetric spaces $\frac{SL(2,\mathbb{R})}{SO(1,1)}$ and $\frac{SL(2,\mathbb{R})}{SO(2)}$ in the vicinity of the identity, before finding simple solutions to the equations of motion. The embedding of the solutions in space-time will be left for the following section.
\subsection{A two-parameter $\sigma$-model on $\frac{SL(2,\mathbb{R})}{SO(1,1)}$}
In this section we will seek a solution on the symmetric space which depends on two parameters and solves the equations of motion of the corresponding $\sigma$-model. We will probe the symmetric space only near the identity so that we adopt the Minkowski metric. { \cb The generalisation of the one-dimensional $\sigma$-model action to a two-dimensional action is a two-dimensional, reparameterisation-invariant $\sigma$-model, which formally describes an auxiliary string worldsheet on the symmetric space.} It is given by
\begin{equation}
S=\int d\sigma d\tau \sqrt{-h} (P_\alpha | P_\beta) h^{\alpha\beta}
\end{equation}
where $h_{\alpha\beta}$ is a metric on the coset $\frac{SL(2,\mathbb{R})}{SO(1,1)}$, $h$ is its determinant and local to the identity element the metric on the coset manifold $\frac{SL(2,\mathbb{R})}{SO(1,1)}$ has orthonormal basis elements $X\equiv \frac{1}{\sqrt{2}}{H}=\frac{1}{\sqrt{2}}\left(\begin{array}{cc}
1 & 0 \\
0 & -1
\end{array}\right)$ and $Y\equiv \frac{1}{\sqrt{2}}{(E-F)}=\frac{1}{\sqrt{2}}\left(\begin{array}{cc}
0 & 1 \\
-1 & 0
\end{array}\right)$, where the Cartan-Killing form is $(A|B)=Tr(AB)$ for two matrices $A$ and $B$ on the manifold. One can check that a vector displaced from the identity element by $x X + y Y$ has length $x^2-y^2$ using the Cartan-Killing form. This is the quadratic form on $\frac{SL(2,\mathbb{R})}{SO(1,1)}$ which is invariant under $SO(1,1)$ transformations. Coordinates $x$ and $y$ parameterise the tangent space of $\frac{SL(2,\mathbb{R})}{SO(1,1)}$ and functions on the symmetric space will be functions of these two variables. The inner product on the tangent space at the identity\footnote{The symmetric space is the single-sheeted hyperboloid and the left-invariant metric is the de Sitter metric $ds^2=(du^2-dv^2)/v^2$, where $u$ and $v$ are functions of $x$ and $y$.} is the Minkowski metric, and in the vicinity of the identity it is approximately the Minkowski metric. In anticipation of identifying $x$ and $y$ with coordinates of $X$ in space-time $\cal M$ we will adopt the notation $\sigma$ and $\tau$, in place of $x$ and $y$, where $\sigma$ and $\tau$ will stand for a pair of space-time coordinates, one of which is space-like and the other time-like. We emphasise that the only constraint that will be placed on the spacetime coordinates is that they parameterise a sub-space of ${\cal M}$ with an $SO(1,1)$ isometry. In identifying the truncation ${\overline{E_{11}}}=SL(2,\mathbb{R})$ there is an associated truncation of space-time coordinates to ${\overline X}$, however we may choose to associate $\sigma$ and $\tau$ with an alternative pair of coordinates in $X$ whose norm is preserved by $SO(1,1)$ transformations; we make use of $\sigma$ and $\tau$ in the following to denote this pair of coordinates in space-time in an abstract manner before discussing particular solutions in section \ref{Mtheories} at which point $\sigma$ and $\tau$ will be identified with space-time coordinates\footnote{With these comments we draw the reader's attention to a general peculiarity of the construction of brane $\sigma$-models namely that $\overline{X}$ will be associated with exotic coordinates, coordinates other than $x^\mu$ of supergravity in eleven dimensions. In order to make contact with known supergravity solutions, the natural $\overline{X}$ is foresaken in favour of $\overline{X}=\{x^\mu\}$. We continue this choice in the present work but note that the conclusion of our work will be that adoption of exotic coordinates in solutions is ultimately unavoidable.} in $X$. Restricting our attention herein to the vicinity of the identity element we take $h_{\alpha\beta}\approx \eta_{\alpha\beta}$ ($\eta_{\alpha\beta}$ being the Minkowski metric). Using the parameterisation of the coset group element $g=\exp(\phi H)\exp(C E)$, where $\phi=\phi(\sigma,\tau)$ and $C=C(\sigma,\tau)$, the action becomes
\begin{equation}
S=\int d\sigma d\tau \sqrt{-h} (2\partial_\alpha \phi \partial^\alpha \phi -\frac{1}{2} e^{4\phi}\partial_\alpha C\partial^\alpha C). \label{so11sigmamodel}
\end{equation}
The equations of motion are
\begin{align}
\partial_\alpha\partial^\alpha\phi+\frac{1}{2}e^{4\phi}\partial_\alpha C\partial^\alpha C&=0 \label{phieom}\\
\partial_\alpha(e^{4\phi}\partial^\alpha C)&=0 \label{Ceom}\\
-\eta_{\alpha\beta}(\partial_\gamma\phi\partial^\gamma\phi -\frac{1}{4}e^{4\phi}\partial_\gamma C \partial^\gamma C)+2\partial_\alpha\phi\partial_\beta\phi-\frac{1}{2}e^{4\phi}\partial_\alpha C \partial_\beta C&=0\label{heom}.
\end{align}
where we have set $h_{\alpha\beta}=\eta_{\alpha\beta}$. 

Equations ({\ref{phieom}-\ref{heom}}) admit simple solutions which depend on only one coset coordinate which we list in cases $(i)-(iv)$ below. In case $(v)$ we exhibit the solution for which both the fields $\phi$ and $C$ depend on both coset coordinates.

\subsubsection{Case (i): $\phi\equiv \phi(\sigma ), C\equiv C(\sigma)$}
The equations $(\ref{phieom}-\ref{heom})$ reduce to those in equation $(\ref{nullgeodesic})$ where the derivative $\partial_\alpha=\partial_\sigma$. As described earlier the solutions of this form, once embedded in space-time and oxidised, include the $\frac{1}{2}$-BPS brane solutions.

\subsubsection{Case (ii): $\phi\equiv \phi(\tau ), C\equiv C(\tau)$} \label{case2}
The equations $(\ref{phieom}-\ref{heom})$ reduce to those in equation $(\ref{nullgeodesic})$ where the derivative $\partial_\alpha=\partial_\tau$. The equations are solved by
\begin{equation}
\phi=\frac{1}{2}\ln(N),\quad C=-N^{-1}+a \label{case2solution}
\end{equation}
where $N=b+q\tau$; $a$, $b$ and $q$ are constants. We will embed a solution of this type into eleven dimensional space-time in the following section.

\subsubsection{Case (iii): $\phi\equiv \phi(\tau ), C\equiv C(\sigma)$} \label{case3}
The equations $(\ref{phieom}-\ref{heom})$ are solved by
\begin{equation}
\phi=-\frac{1}{2}\ln(N),\quad C=a+q\sigma
\end{equation}
where $N=b+q\tau$; $a$, $b$ and $q$ are constants.
This solution relies upon the linearity of $N$ and $C$ in the coordinates $\sigma$ and $\tau$ respectively and is a solution only in two dimensions. To be convinced of this, consider embedding $(\ref{phieom}-\ref{heom})$ in $\mathbb{R}^{2,1}$ with $x^\alpha\in \{\tau,\sigma_1,\sigma_2\}$ and seeking solutions such that $C\equiv C(\sigma_1,\sigma_2)$ and $\phi\equiv\phi(\tau)$. Equation $(\ref{Ceom})$ implies that $\partial_i\partial_i C=0$ where $x^i\in \{\sigma_1,\sigma_2\}$ while equation $(\ref{phieom})$ implies that $(\partial_i C)(\partial_i C)$ is a constant - giving $C=a+\vec{k}\cdot\vec{\sigma}$ where $\vec{k}\cdot\vec{\sigma}=k_1\sigma_1+k_2\sigma_2$ and $k_1$, $k_2$ are constants. Setting $\phi=-\frac{1}{2}\ln(N)$ where $N=b+|\vec{k}|\tau$ gives a solution to both equations $(\ref{phieom})$ and $(\ref{Ceom})$. However equation $(\ref{heom})$ constrains the solution to be intrinsically two-dimensional, the $\alpha=2$, $\beta=3$ equation reduces to $\partial_2 C\partial_3C=k_1k_2=0$, requiring $C$ to be a function of only one of the two spatial coordinates. Furthermore if we set $k_2=0$, the $\alpha=3$, $\beta=3$ equation is only solved if $k_1=0$ too. This solution is intrinsic to its embedding in a two-dimensional space-time and does not have a corresponding solution in eleven dimensions.

\subsubsection{Case (iv): $\phi\equiv \phi(\sigma ), C\equiv C(\tau)$}
The equations $(\ref{phieom}-\ref{heom})$ are solved by
\begin{equation}
\phi=-\frac{1}{2}\ln(N),\quad C=a+q\tau
\end{equation}
where $N=1+b\sigma$, $a$ and $q$ are constants and the form of $N$ is fixed by assuming the background space-time is Minkowski in the limit $\sigma \rightarrow 0$. As for case (iii) above this solution is intrinsic to a two-dimensional space-time and does not have a corresponding solution in eleven dimensions.

\subsubsection{Case (v): $\phi\equiv \phi(\sigma,\tau ), C\equiv C(\sigma,\tau)$}\label{case5}
Let\footnote{We adopt a convention for solving equation (\ref{Ceom}) that aims to use a similar notation as used for solving the one-parameter model in section \ref{oneparameterequations}. A standard approach to solving equation (\ref{Ceom}) is to write $e^{4\phi}\partial^\alpha C=\epsilon^{\alpha\beta\gamma}\partial_\beta V_\gamma,$ where $V_\gamma$ is a function of $\tau$ and $\sigma$, which solves the equation of motion. The notation we have adopted is related to $V_3$ by $\partial^\alpha A = \epsilon^{\alpha\beta\gamma}\partial_\beta V_\gamma$.} $\partial^\alpha A\equiv e^{4\phi}\partial^\alpha C$ then equation (\ref{Ceom}) reduces to the wave equation in one spatial dimension for $A$:
\begin{equation}
\partial_\sigma^2 A - \partial_\tau^2 A =0
\end{equation}
and $A=f(\sigma+\tau)+g(\sigma-\tau)$, where $f$ and $g$ are arbitrary functions. Rewriting $\phi(\sigma,\tau)\equiv \frac{1}{2}\ln(u(\sigma+\tau)+v(\sigma-\tau))$ and substituting this and the expression for $A$ into equation (\ref{phieom}) gives
\begin{equation}
f'g'=u'v'
\end{equation}
where a prime denotes a derivative with respect to the argument. 
There are three independent equations contained in equation (\ref{heom}):
\begin{align}
-\partial_\gamma\phi\partial^\gamma\phi +\frac{1}{4}e^{4\phi}\partial_\gamma C \partial^\gamma C+2\partial_\sigma\phi\partial_\sigma\phi-\frac{1}{2}e^{4\phi}\partial_\sigma C \partial_\sigma C&=0\label{heomsigsig}\\
\partial_\gamma\phi\partial^\gamma\phi -\frac{1}{4}e^{4\phi}\partial_\gamma C \partial^\gamma C+2\partial_\tau\phi\partial_\tau\phi-\frac{1}{2}e^{4\phi}\partial_\tau C \partial_\tau C&=0\label{heomtautau}\\
2\partial_\sigma\phi\partial_\tau\phi-\frac{1}{2}e^{4\phi}\partial_\sigma C \partial_\tau C&=0\label{heomsigtau}.
\end{align}
Addition of equations (\ref{heomsigsig}) and (\ref{heomtautau}) yields
\begin{equation}
2\partial_\sigma\phi\partial_\sigma\phi+2\partial_\tau\phi\partial_\tau\phi-\frac{1}{2}e^{4\phi}(\partial_\sigma C \partial_\sigma C+\partial_\tau C \partial_\tau C)=0
\end{equation}
while their difference is trivial. Rewriting the equation above in terms of $u$, $v$, $f$ and $g$ gives
\begin{equation}
(u')^2+(v')^2=(f')^2+(g')^2.
\end{equation}
Hence a simple solution is described by the two travelling wave functions $f(\sigma+\tau)$ and $g(\sigma-\tau)$, i.e. by the fields
\begin{equation}
\phi=\frac{1}{2}\ln(f+g)\qquad \mbox{and} \qquad C=-\frac{1}{f+g}.
\end{equation}

\subsection{No non-trivial two-parameter solutions to the brane $\sigma$-model on $\frac{SL(2,\mathbb{R})}{SO(2)}$}
Let 
\begin{equation}
S_E=\int d\sigma_1 d\sigma_2 \sqrt{h} (P_\alpha | P_\beta) h^{\alpha\beta} \label{so2sigmamodel}
\end{equation}
where $h_{\alpha\beta}$ is a metric\footnote{The space is the hyperbolic plane, and the metric is the Poincar\'e half-plane metric $ds^2=(du^2+dv^2)/v^2$, where $u$ and $v$ are functions of $x$ and $y$.} on the coset $\frac{SL(2,\mathbb{R})}{SO(2)}$, and we will in the vicinity of the identity element and set $h_{\alpha\beta}=\delta_{\alpha\beta}$. 
The action becomes
\begin{equation}
S_E=\int d\sigma_1 d\sigma_2 \sqrt{h} (2\partial_\alpha \phi \partial^\alpha \phi +\frac{1}{2} e^{4\phi}\partial_\alpha C\partial^\alpha C).
\end{equation}
The equations of motion are
\begin{align}
\partial_\alpha\partial^\alpha\phi-\frac{1}{2}e^{4\phi}\partial_\alpha C\partial^\alpha C&=0 \label{phieom2}\\
\partial_\alpha(e^{4\phi}\partial^\alpha C)&=0 \label{Ceom2}\\
-\delta_{\alpha\beta}(\partial_\gamma\phi\partial^\gamma\phi +\frac{1}{4}e^{4\phi}\partial_\gamma C \partial^\gamma C)+2\partial_\alpha\phi\partial_\beta\phi+\frac{1}{2}e^{4\phi}\partial_\alpha C \partial_\beta C&=0\label{heom2}
\end{align}
where we have set $h_{\alpha\beta}=\delta_{\alpha\beta}$.

Let $\partial^\alpha B\equiv e^{4\phi}\partial^\alpha C$ then equation (\ref{Ceom2}) reduces to the Laplace equation in two dimensions for $B(\sigma_1,\sigma_2)$:
\begin{equation}
\partial_{\sigma_1}^2 B + \partial_{\sigma_2}^2 B =0
\end{equation}
and $B=\hat{f}(\sigma_1+i\sigma_2)+\hat{g}(\sigma_1-i\sigma_2)$ is a real function.
Rewriting $\phi(\sigma_1,\sigma_2)\equiv \frac{1}{2}\ln(\hat{u}(\sigma_1+i\sigma_2)+\hat{v}(\sigma_1-i\sigma_2))$, such that $\phi$ remains a real function, and substituting this and the expression for $B$ into equation (\ref{phieom2}) gives
\begin{equation}
\hat{u}'\hat{v}'=\hat{f}'\hat{g}'
\end{equation}
where a prime denotes a derivative with respect to the argument. 
There are three independent equations contained in equation (\ref{heom2}):
\begin{align}
-\partial_\gamma\phi\partial^\gamma\phi -\frac{1}{4}e^{4\phi}\partial_\gamma C \partial^\gamma C+2\partial_{\sigma_i}\phi\partial_{\sigma_i}\phi+\frac{1}{2}e^{4\phi}\partial_{\sigma_i} C \partial_{\sigma_i} C&=0 \quad \mbox{for } i\in\{1,2\}\label{heomsigisigi}\\
2\partial_{\sigma_1}\phi\partial_{\sigma_2}\phi+\frac{1}{2}e^{4\phi}\partial_{\sigma_1} C \partial_{\sigma_2} C&=0\label{heomsig1sig2}.
\end{align}
Subtracting the $i=2$ equation of (\ref{heomsigisigi}) from the $i=1$ equations yields
\begin{equation}
(\hat{u}')^2+(\hat{v}')^2=-((\hat{f}')^2+(\hat{g}')^2)
\end{equation}
while their sum is trivial. Hence we have
\begin{equation}
((e^{2\phi})')^2=(\hat{u}'+\hat{v}')^2=-(\hat{f}'+\hat{g}')^2=-(B')^2
\end{equation}
which has no non-trivial real solutions for $B(\sigma_1,\sigma_2)$ and $\phi(\sigma_1,\sigma_2)$. 

There do exist some simple solutions to these equations which are of the form $\phi\equiv \phi(\sigma_1 ), C\equiv C(\sigma_2)$. The equations (\ref{phieom2} - \ref{heom2}) are then solved by
\begin{equation}
\phi=-\frac{1}{2}\ln(N),\quad C=a+q\sigma_2 \label{so2solution}
\end{equation}
where $N=b+q\sigma_1$; $a$, $b$ and $q$ are constants. However these solutions are intrinsic to two dimensions and do not admit an embedding in higher dimensions for reasons similar to those given earlier in section \ref{case3}.

\section{$M$-theory, $M^*$-theory and $M'$-theory Solutions}\label{Mtheories}
The solutions found in the preceding section are not simple to embed in the eleven-dimensional space-time of $M$-theory. The local Lorentz group of $M$-theory is $SO(1,10)$; space-time consists of a single temporal coordinate and ten spatial coordinates. The symmetric space $\frac{SL(2,\mathbb{R})}{SO(1,1)}$ is a non-compact, pseudo-Riemannian manifold and any map from this manifold into a two-dimensional sub-space of space-time, transverse to the world volume of the space-time solution\footnote{In the simple cases this will be a p-brane, which splits the isometries of space-time into the product $SO(D-p-1)\times SO(1,p)$, where the $SO(1,p)$ isometries act on the world-volume of the brane and the $SO(D-p-1)$ isometries act on the transverse space. For more complicated solutions there world-volume isometries will be further split, but the notion of transverse space remains well-defined, and may be inferred from the root of $E_{11}$ used to construct the coset as a truncation of $E_{11}$.} must preserve the $SO(1,1)$ isometries. This presents some immediate problems in applying this method to the standard (electric) branes of $M$-theory. Consider the $M2$-brane: to construct the coset $\frac{SL(2,\mathbb{R})}{SO(1,1)}$, the three-form generator $E_\alpha$ must transform under the involution $\Omega(E_\alpha)=F_{\alpha}$. Consistency under Poincar\'e duality implies that the $M2$-brane world-volume must have an odd number of temporal coordinates on its world-volume and consistency with $M$-theory means there is only a single temporal direction in space-time and that it lies on the world-volume of the brane. In short, the transverse space for standard (electric $\frac{1}{2}$-BPS branes) $M$-theory solutions is a Riemannian manifold. How might one introduce a pseudo-Riemannian transverse space? In the context of $E_{11}$ there is the possibility to consider the $M^*$ and $M'$-theories \cite{Hull:1998fh} which have two and five temporal coordinates respectively\footnote{The exotic signatures $(2,9)$, $(5,6)$, $(6,5)$ and $(9,2)$ were first understood to be relevant to M-theory in \cite{Blencowe:1988sk}.}. The solutions presented in the previous section will be embedded into both $M^*$ and $M'$- theories, for cases where the transverse space admits a two-dimensional sub-space with $SO(1,1)$ isometry. Both $M^*$-theory and $M'$-theory are consistent with an $E_{11}$ symmetry of $M$-theory; they correspond to particular Weyl reflections of $M$-theory solutions. Consequently solutions in $M^*$ and $M'$-theories that we construct will be related by an $E_{11}$ Weyl reflection to sectors of $M$-theory. In this section we will first present the embedding in space-time of a set of particular solutions to the two-parameter $\sigma$-model described earlier, before presenting a method for embedding the most general solutions we have found in $M^*$ and $M'$-theories. Finally we will investigate the possibility of Weyl-reflecting $M^*$ and $M'$ solutions to $M$-theory.

\subsection{Particular solutions}
In section \ref{2dsigmamodel} we constructed solutions to the equations of motion of (\ref{so11sigmamodel}) and (\ref{so2sigmamodel}). The special solutions to the $\sigma$-model on $\frac{SL(2,\mathbb{R})}{SO(1,1)}$, discussed case-by-case in section \ref{case2}, will be investigated here.

\subsubsection{Cosmological Collapsing Solutions in $M^*$-theory} 
The solution to the $\sigma$-model defined on $\frac{SL(2,\mathbb{R})}{SO(1,1)}$ in section \ref{case2} is given in equation (\ref{case2solution}). The fields depend only on the temporal coordinate\footnote{Alternative approaches to constructing a vast range of cosmological solutions and extremal $S$-branes from the one-parameter $\sigma$-model have been studied in \cite{Kleinschmidt:2005gz}.}: $\phi\equiv\phi(\tau)$ and $C\equiv C(\tau)$. In this example we identify the global symmetry group of the $\sigma$-model with the truncation of $E_{11}$ to $SL(2,\mathbb{R})$ given in (\ref{M2embedding1}) and (\ref{M2embedding2}). Compared to the reconstruction of the M2-brane, via the supergravity dictionary, described in section \ref{M2brane} we now expect two-parameter solutions to have an $SO(1,1)$ isometry in a subspace transverse to the brane, i.e. these solutions require there to be at least one temporal coordinate and one spatial coordinate transverse to the brane. This example is a special case, as the solution depends only on the temporal coordinate on the symmetric space, hence we require the signature of space-time to have a time coordinate transverse to the brane. In addition we require the signature function to correspond to a temporal involution $\Omega$ which picks out ${\cal K}(G)=SO(1,1)$ when the $E_{11}$ algebra is truncated to $E_\beta$, $H_\beta$ and $F_\beta$. The action of the temporal involution and the signature function are related by
\begin{equation}
\Omega(E_\beta)=-\epsilon_\beta F_\beta	\qquad \mbox{where}\qquad \epsilon_\beta =(-1)^{<\vec{\beta},\vec{f}>}.
\end{equation}
Hence $E_\beta - (-1)^{<\vec{\beta},\vec{f}>}F_\beta \in \mathfrak{so}(1,1)$ when $<\vec{\beta},\vec{f}>=1\mod 2$. Additionally we require that $i_0(\vec{F})=1\mod 2$, where $i_0$ is defined in equation (\ref{i0}) to guarantee Poincar\'e duality \cite{Keurentjes:2005jw} and furthermore that the signature function $\vec{f}$ is in the $E_{11}$ Weyl orbit of the $M$-theory signature function $\vec{f}=\lambda_1$. These conditions constrain the roots $\vec{\beta}$ and signatures for which the solutions may be embedded in space-time.

\underline{\bf {The M2-root: $\vec{\beta}=\vec{\alpha_{11}}$ in background signature $(2,9)$}.} There are three classes of signature function which may be distinguished by the number of temporal directions among the brane world-volume coordinates $\{x^9,x^{10},x^{11}\}$. The signature functions which lie in the Weyl orbit of the $M$-theory signature function $\vec{f}=\vec{\lambda_1}$ and for which $\epsilon_\beta=-1$ requires the pair of temporal coordinates to be either both longitudinal to the brane (e.g. $\vec{f}=\vec{\lambda_8}+\vec{\lambda_{10}}+\vec{\lambda_{11}}$, where $x^9$ and $x^{10}$ are temporal coordinates so that $\epsilon_0=0$) or both transverse to the brane (e.g. $f=\vec{\lambda_2}+\vec{\lambda_{11}}$ where $x^1$ and $x^2$ are timelike coordinates, so that $\epsilon_0=1$). As we require that the transverse space contain a temporal coordinate we take our signature function from the second example. Let us identify the coset model parameters with space-time coordinates according to $\tau=x^1$, a temporal coordinate transverse to the brane. The $\phi$ and $C$ fields of the coset will be associated with the elfbein and the membrane gauge field $A_{91011}$ to give
\begin{align}
{e_{\hat{1}}}^1&={e_{\hat{2}}}^2=\ldots = {e_{\hat{8}}}^8=e^{\frac{1}{6}\ln(N)}=N^{\frac{1}{6}}, \label{M2elfbein1}\\
{e_{\hat{9}}}^9&={e_{\hat{10}}}^{10}={e_{\hat{11}}}^{11}=e^{-\frac{1}{3}\ln(N)}=N^{-\frac{1}{3}} \label{M2elfbein2}\\
F_{\hat{1}\hat{9}\hat{10}\hat{11}}&=N^{-2}\partial_{\hat{1}}(N)
\end{align}
where $N=b+qx^1$ and the hatted index denotes a space-time index (the field strength components have been constructed from the dictionary $P_{\alpha}=e^{2\phi}\partial_\alpha C\equiv F_{\alpha 9 10 11}$). The space-time metric is 
\begin{equation}
ds^2=N^{1/3}(-(dx^1)^2-(dx^2)^2+\ldots +(dx^8)^2)+N^{-2/3}((dx^9)^2+(dx^{10})^2+(dx^{11})^2)
\end{equation}
which is a solution to the equations of motion derived from the eleven-dimensional action
\begin{equation}
S_{M^*}=\int R\star 1 + \frac{1}{2}F\wedge \star F \label{M2*action}
\end{equation}
with space-time signature $(2,9)$, obtained by substituting $\vec{f}=\vec{\lambda_2}+\vec{\lambda_{11}}$ and $\epsilon_0=1$ (so that both $x^1$ and $x^2$ are temporal coordinates) into the first terms of equation (\ref{genericlowlevelaction}). We observe that, as $\tau=x^1$ (one of the two time coordinates) evolves, $1/N$ is suppressed, which, in terms of the metric, corresponds to the shrinking of the three-dimensional brane world-volume, and the expansion of an eight-dimensional space-time with symmetry $SO(2,6)$. While this results in an emergent space-time far from the physical universe, the process through which part of the eleven dimensional space-time collapses may be interesting.
Examples using other low-level roots of $E_{11}$ follow a similar path: a root associated with the $M5$ brane solution gives rise to a space-time with an $SO(2,3)$ isometry, with a shrinking six-dimensional space as the second time coordinate evolves. The construction associated with the dual elfbein at level 3 in the decomposition of $E_{11}$ is more slightly more involved and of interest:

\underline{\bf {The dual elfbein root in background signature $(2,9)$}.} The relevant symmetric space is constructed by taking the root 
$\vec{\beta}=\vec{\alpha_{4}}+2\vec{\alpha_{5}}+3\vec{\alpha_{6}}+4\vec{\alpha_{7}}+5\vec{\alpha_{8}}+3\vec{\alpha_{9}}+\vec{\alpha_{10}}+3\vec{\alpha_{11}}$ as the single real positive root of the root system of $SL(2,\mathbb{R})$. The $SL(2,\mathbb{R})$ generators are
\begin{align}
H&=-({K^1}_1+{K^2}_2+{K^3}_3)+{K^{11}}_{11},\\ 
E&=R^{4567891011|11} \mbox{ and } F=R_{4567891011|11}.
\end{align}
There are five classes of signature function to consider which may be distinguished by the number of temporal directions among the coordinates $\{x^1,x^2,x^3\}$, $\{x^4,x^5,\ldots , x^{10}\}$ and $\{x^{11}\}$. Of these only two classes of signature function lie in the Weyl orbit of the $M$-theory signature function $\vec{f}=\vec{\lambda_1}$ and satisfy $\epsilon_\beta=(-1)^{<\vec{\beta},\vec{f}>}=-1$. The first signature function requires one of the temporal coordinates to be $x^{11}$ and the other to be one of the coordinates $\{x^1,x^2,x^3\}$ (e.g. $\vec{f_1}=\vec{\lambda_1}+\vec{\lambda_{10}}$ where $x^1$ and $x^{11}$ are temporal coordinates if $\epsilon_0=1$). The second possible class of signature function contain both temporal coordinates in the set $\{x^1,x^2,x^3\}$ (e.g. $\vec{f_2}=\vec{\lambda_2}+\vec{\lambda_{11}}$ where $x^1$ and $x^2$ are timelike coordinates if $\epsilon_0=1$). In this example the coordinates $\{x^1,x^2,x^3\}$ form the ``transverse" space so we may consider both signature functions $\vec{f_1}$ and $\vec{f_2}$ defined above. 

For the first signature function $\vec{f_1}=\vec{\lambda_1}+\vec{\lambda_{10}}$ (where $x^1$ and $x^{11}$ are temporal coordinates) we identify the coset model parameters with space-time coordinates by $\tau=x^1$. The $\phi$ and $C$ fields of the $\sigma$-model are associated with the elfbein and the dual elfbein gauge field $A_{4567891011|11}$ which is dual to an off-diagonal component of the metric:
\begin{align}
{e_{\hat{1}}}^1&={e_{\hat{2}}}^2 = {e_{\hat{3}}}^3=e^{\frac{1}{2}\ln(N)}=N^{\frac{1}{2}},\label{TNelfbein1}\\
{e_{\hat{4}}}^4&={e_{\hat{5}}}^5 =\ldots = {e_{\hat{10}}}^{10}=1\\
{e_{\hat{11}}}^{11}&=N^{-\frac{1}{2}}\\
{e_{\hat{2}}}^{11}&=\frac{1}{2}q x^3 N^{-\frac{1}{2}} \qquad \mbox{ and } \qquad {e_{\hat{3}}}^{11}=-\frac{1}{2}q x^2 N^{-\frac{1}{2}}\label{TNelfbein4}
\end{align}
where $N=b+qx^1$. The space-time metric is 
\begin{align}
ds^2&=Nd\Sigma_{(1,2)}^2+d\Omega_7^2-N^{-1}(dx^{11}-\frac{1}{2}q x^3 dx^2+\frac{1}{2}q x^2 dx^3)^2
\end{align}
which is a solution of the vacuum Einstein equations derived from varying the action given in equation \eqref{M2*action} where $x^1$ and $x^{11}$ are temporal coordinates\footnote{The sign of the kinetic term for the dual elfbein field strength is positive, i.e. $+\frac{1}{2}F\wedge \star F$ appears in the action. The interested reader is referred to footnote \ref{footnote_on_signs} to see how this sign is determined from the signature function $\vec{f}$.}. 
The temporal coordinate $x^1$ interpolates between $\mathbb{R}^{2,9}$ when $x^1=0$ and $\mathbb{R}^{1,2}\times \mathbb{R}^7$ when $x^1\to \infty$, where due to the evolution of the solution under one temporal coordinate $x^1$ we find that the second temporal coordinate $x^{11}$ is suppressed.

The solution above corresponds to a one-dimensional version\footnote{The harmonic function of the solution depends on only $x^1$, rather than on $x^1$, $x^2$ and $x^3$.} of the Taub-NUT solution in eleven dimensions with two time-coordinates. It can be unsmeared to give one version of the Taub-NUT solution in a background with two times (the second version, which is derived from the alternative signature function, will be given below):
\begin{equation}
ds^2=N(dr^2-r^2d\phi^2+r^2\cosh^2{\phi}d\theta^2)-N^{-1}(dx^{11}-q\sinh{\phi}d\phi)^2+d\Omega_7^2	
\end{equation}
where $N=1+\frac{q}{r}$ and we have changed to (single-sheeted) hyperbolic coordinates according to $x^1=r\sinh{\phi}$, $x^2=r\cosh{\phi}\cos{\theta}$ and $x^3=r\cosh{\phi}\sin{\theta}$, so that $r^2=-(x^1)^2+(x^2)^2+(x^3)^2$. To remove the conical singularity apparent as $\phi\to 0$ and $r\to 0$, $\theta$ has period $4\pi$. 

For the second signature function $\vec{f_2}=\vec{\lambda_2}+\vec{\lambda_{11}}$ (where $x^1$ and $x^2$ are temporal coordinates) we again identify the coset model parameter in the solution with space-time coordinates by $\tau=x^1$. The non-zero elfbein components are the same as those in equations (\ref{TNelfbein1}-\ref{TNelfbein4}), but due to the change of signature the space-time metric is altered to
\begin{align}
ds^2&=Nd\Sigma_{(2,1)}^2+d\Omega_7^2+N^{-1}(dx^{11}-\frac{1}{2}q x^3 dx^2+\frac{1}{2}q x^2 dx^3)^2
\end{align}
which is a solution of the vacuum Einstein equations in the $(2,9)$ signature where $x^1$ and $x^2$ are the temporal coordinates. 
The temporal coordinate $x^1$ interpolates between $\mathbb{R}^{(2,9)}$ when $x^1=0$ and $\mathbb{R}^{2,1}\times \mathbb{R}^7$ when $x^1\to \infty$. This one-dimensional solution of the vacuum equations can be unsmeared to give a second type of Taub-NUT in a space-time with two temporal coordinates:
\begin{equation}
ds^2=N(dr^2-r^2d\phi^2-r^2\sinh^2{\phi}d\theta^2)+N^{-1}(dx^{11}-q\cosh{\phi}d\phi)^2+d\Omega_7^2	
\end{equation}
where $N=1+\frac{q}{r}$ and we have changed to (two-sheeted) hyperbolic coordinates according to $x^1=r\sinh{\phi}\cos{\theta}$, $x^2=r\sinh{\phi}\sin{\theta}$ and $x^3=r\cosh{\phi}$, so that $r^2=-(x^1)^2-(x^2)^2+(x^3)^2$. When $r\to \infty$ the solution is asymptotically locally flat. To remove the conical singularity apparent as $r\to 0$, $\theta$ has period $4\pi$.

\subsection{General solutions} 
In this section we will embed in space-time the two-parameter $\sigma$-model solutions given in case (v) in section \ref{case5} in which both fields of the $\sigma$-model depend on $\sigma$ and $\tau$. We will restrict our attention to the example of the membrane in $M^*$-theory and in $M'$-theory, although generalisations of the pp-wave, the five-brane, the KK6-brane and other exotic $E_{11}$ branes, as well as bound states, may be constructed in this way. The membrane solutions constructed from the two-parameter $\sigma$-model reproduce the membrane solutions with wordlvolume signature $(3,0)$ and $(1,2)$ found in \cite{Hull:1998fh}, which justifies the construction outlined in this paper. The principal new feature of the solutions is that they are defined in terms of wavefunctions rather than harmonic functions. 

\label{M2*brane}
\subsubsection{The $M2^*$ brane}
Let the $\frac{SL(2,\mathbb{R})}{SO(1,1)}$ coset be defined by the $E_{11}$ generators associated with the supergravity membrane as given in equations (\ref{M2embedding1}) and (\ref{M2embedding2}). As described earlier we will be interested in the signature function $\vec{f}=\vec{\lambda_2}+\vec{\lambda_{11}}$ (where $x^1$ and $x^2$ are the two temporal coordinates and are transverse to the membrane's world-volume). For the embedding we identify the coset coordinates $(\sigma,\tau)$ with space-time coordinates transverse to the brane world-volume e.g. $\tau=x^1$, $\sigma=x^3$. The non-zero elfbein components have the same form as given in equations (\ref{M2elfbein1}) and (\ref{M2elfbein2}), but where now $N=f(x^1+x^3)+g(x^1-x^3)$. The non-zero field strength components are 
\begin{align}
F_{\hat{1}\hat{9}\hat{10}\hat{11}}&=\frac{f'+g'}{(f+g)^2}\\
F_{\hat{3}\hat{9}\hat{10}\hat{11}}&=\frac{f'-g'}{(f+g)^2}.
\end{align}
The full, unsmeared solution, which respects the $SO(2,6)$ isometry of the transverse space is found by modifying the wavefunctions $f$ and $g$ to be $f(\vec{x}\cdot\vec{k}+\vec{\omega}\cdot \vec{t})$ and $g(\vec{x}\cdot\vec{k}-\vec{\omega}\cdot \vec{t})$, where $\vec{\omega}=(\omega_1,\omega_2)^T$, $\vec{k}=(k_1,k_2,\ldots, k_6)^T$, $\vec{t}=(x^1,x^2)^T$ and $\vec{x}=(x^3,x^4,\ldots, x^8)^T$. The dispersion relation is $\vec{\omega}^2={\vec{k}}^2$. The space-time metric is 
\begin{align}
ds^2=&(f+g)^{1/3}(-(dx^1)^2-(dx^2)^2+(dx^3)^2\ldots +(dx^8)^2)+ \nonumber\\
& (f+g)^{-2/3}((dx^9)^2+(dx^{10})^2+(dx^{11})^2) \label{M2*brane1}
\end{align}
and the non-zero components of the four-form are 
\begin{equation}
F_{\hat{\mu} \hat{9}\hat{10}\hat{11}}=\partial_{\hat{\mu}} (\frac{-1}{f+g}) \label{M2*brane2}
\end{equation}
and its antisymmetrisations. The metric and field strength given in equations \eqref{M2*brane1} and \eqref{M2*brane2} solve the equations of the bosonic $M^*$-theory action given in equation \eqref{M2*action}. The special case where $f+g=1+\frac{q}{r^6}$ with $r^2=-(x^1)^2-(x^2)^2+(x^3)^2+\ldots +(x^8)^2$ reproduces the analogue of the $M2$ brane in $M^*$-theory found in \cite{Hull:1998fh}. 

\subsubsection{The $M2'$ brane}
The embedding of the two-parameter $\sigma$-model solution in space-time that gives the $M2'$ brane proceeds in the same manner as for the $M2^*$ brane above. The main difference is the choice of signature function. There are now four distinct classes of signature function, distinguished by the signature on the membrane which may be $(3,0)$, $(2,1)$, $(1,2)$ or $(0,3)$. Subject to the constraints that the signature function lies in the Weyl orbit of the M-theory class of signature functions, that Poincar\'e duality is satisfied $i_0(\vec{f})=1\mod{2}$ and that ${\cal K}(G)=SO(1,1)$ for $\alpha_{11}$ (i.e. that $\epsilon_{11}=(-1)^{<\vec{f},\vec{\alpha_{11}}>}=-1$), only two of the classes of signature functions remain: those with an odd number of time directions on the brane. The representative signature functions from these classes that we will use are $\vec{f'_1}=\vec\lambda_6+\vec\lambda_{11}$ (for which $\{x^7,x^8,x^9,x^{10},x^{11}\}$ are the five time-like coordinates)
and $\vec{f'_2}=\vec\lambda_4+\vec\lambda_9+\vec\lambda_{11}$ (for which $\{x^5,x^6,x^7,x^8,x^9\}$ are the five time-like coordinates). In both cases $\epsilon_0=0$ and the form of the eleven-dimensional action determined from \eqref{genericlowlevelaction} is that of bosonic $M'$-theory:
\begin{equation}
S_{M'}=\int R\star 1 - \frac{1}{2}F\wedge \star F. \label{M2'action}
\end{equation}
Compared to the $M2^*$-brane considered above, there is a difference; here both the signature functions permit a transverse space with an $SO(1,1)$ isometry, so that we find two types of $M2'$-solution. In both cases the $\frac{SL(2,\mathbb{R})}{SO(1,1)}$ coset is defined by the $E_{11}$ generators associated with the supergravity membrane as given in equations (\ref{M2embedding1}) and (\ref{M2embedding2}).

\underline{\bf {Case (i) The $M2'$-brane with world-volume signature $(3,0)$}.} A representative signature function is $\vec{f'_1}=\vec\lambda_6+\vec\lambda_{11}$ and $(\sigma,\tau)$ are identified with space-time coordinates by $\tau=x^7$, $\sigma=x^6$, for example, before unsmearing so that the solution carries the $SO(2,6)$ isometries. The non-zero elfbein components have the same form as given in equations (\ref{M2elfbein1}) and (\ref{M2elfbein2}) and the non-trivial field strength components take the same form as in \eqref{M2*brane2}, but where now $N=f(\vec{x}\cdot\vec{k}+\vec{\omega}\cdot \vec{t})+g(\vec{x}\cdot\vec{k}-\vec{\omega}\cdot \vec{t})$, where $\vec{\omega}=(\omega_1,\omega_2)^T$, $\vec{k}=(k_1,k_2,\ldots, k_6)^T$, $\vec{t}=(x^7,x^8)^T$ and $\vec{x}=(x^1,x^2,\ldots, x^6)^T$. The dispersion relation is $\vec{\omega}^2={\vec{k}}^2$ and the space-time metric is 
\begin{align}
ds^2=&(f+g)^{1/3}((dx^1)^2+(dx^2)^2+\ldots +(dx^6)^2-(dx^7)^2-(dx^8)^2)+ \nonumber\\
& (f+g)^{-2/3}(-(dx^9)^2-(dx^{10})^2-(dx^{11})^2). \label{M2'brane1}
\end{align}

\underline{\bf {Case (ii) The $M2'$-brane with world-volume signature $(1,2)$}}. A representative signature function is $\vec{f'_2}=\vec\lambda_4+\vec\lambda_9+\vec\lambda_{11}$ and $(\sigma,\tau)$ are identified with space-time coordinates by $\tau=x^4$, $\sigma=x^5$, for example, before unsmearing so that the solution carries the $SO(4,4)$ isometries. The non-zero elfbein components have the same form as given in equations (\ref{M2elfbein1}) and (\ref{M2elfbein2}) and the non-trivial field strength components take the same form as in \eqref{M2*brane2}, but where now $N=f(\vec{x}\cdot\vec{k}+\vec{\omega}\cdot \vec{t})+g(\vec{x}\cdot\vec{k}-\vec{\omega}\cdot \vec{t})$, where $\vec{\omega}=(\omega_1,\omega_2,\omega_3,\omega_4)^T$, $\vec{k}=(k_1,k_2,k_3,k_4)^T$, $\vec{t}=(x^5,x^6,x^7,x^8)^T$ and $\vec{x}=(x^1,x^2,x^3,x^4)^T$. The dispersion relation is $\vec{\omega}^2={\vec{k}}^2$ and the space-time metric is 
\begin{align}
ds^2=&(f+g)^{1/3}((dx^1)^2+\ldots+(dx^4)^2-(dx^5)^2-\ldots-(dx^8)^2)+ \nonumber\\
& (f+g)^{-2/3}(-(dx^9)^2+(dx^{10})^2+(dx^{11})^2). \label{M2'brane2}
\end{align}

\subsection{Mapping $M^*$ and $M'$-theory solutions to $M$-theory.}
In order to construct a two-parameter solution from a symmetric space it was necessary to embed the solution in a multiple-time space-time, where at least one temporal coordinate was transverse to the brane. In $M$-theory electric-brane solutions are defined in terms of harmonic functions solving the Laplace equation in the coordinates transverse to the brane, while, in backgrounds with multiple time-coordinates, solutions are defined in terms of wavefunctions (i.e. the Laplace equation is modified to a wave equation when there are temporal transverse coordinates). Solutions of the $\sigma$-models that we have considered are preserved by Weyl reflections. Consider a Weyl reflection, $S_\beta$, a reflection in the plane perpendicular to a root $\vec{\beta}$, it acts on a group element $g$ by $g\to U_{\beta}\, g \, U_{\beta}^{-1}$ where $U_\beta = \exp(F_\beta)\exp(-E_\beta)\exp(F_\beta)$. For truncations of $E_{11}$ to finite matrix subgroups, this transformation of a group element $g$ is straightforward to compute. Note that the Weyl reflection leaves the brane $\sigma$-model invariant as $S_\beta(\nu_\mu)=S_\beta(\partial_\mu g \,g^{-1})=U_\beta \nu_\mu U_\beta^{-1}$ and hence $(\nu | \nu)$ is invariant. Consequently, if a group element encodes a solution of the brane $\sigma$-model, then so does its Weyl reflection. 

The Weyl reflections of $E_{11}$ do not preserve the signature of the background space-time, while the Weyl reflections do map solutions in $M^*$ and $M'$-theory to solutions in $M$-theory. Consequently the solutions found in section $\ref{M2*brane}$, which are parameterised by arbitrary travelling wave functions $f(\vec{k}\cdot\vec{x}+\vec{\omega}\cdot\vec{t})$ and $g(\vec{k}\cdot\vec{x}-\vec{\omega}\cdot\vec{t})$, are mapped under the appropriate Weyl reflections to a solution in $M$-theory. The action of the Weyl reflections is to map one choice of an $SL(2,\mathbb{R})$ sub-group in $E_{11}$ to another but it does not change the wavefunctions apparent in the $M^*$ and $M'$-theory solutions. It is natural to wonder where these solutions are mapped to in $M$-theory, since the known brane solutions are not dressed with travelling wave functions. 

Let us focus on our prototype solution of the $M2^*$ brane given in equations (\ref{M2*brane1}) and (\ref{M2*brane2}). The signature function was $\vec{f}=\vec{\lambda_2}+\vec{\lambda_{11}}$ and we observe that the Weyl reflection $S_{\beta_{129}}$ where \begin{equation}
\vec{\beta}_{129}=\vec{e_1}+\vec{e_2}+\vec{e_9}=\vec{\alpha_1}+2(\vec{\alpha_2}+\ldots + \vec{\alpha_9})+\vec{\alpha_{10}}+\vec{\alpha_{11}} 	
\end{equation}
maps $\vec{f}$ to 
\begin{equation}
\vec{f}'\equiv \vec{\beta}_{129}(\vec{f})=\vec{\lambda_8}-\vec{\lambda_9}+\vec{\lambda_{11}}	
\end{equation}
which corresponds to an $M$-theory background space-time where $x^9$ is the sole temporal coordinate. Hence we may apply this Weyl reflection to the $M2^*$ solution given in equation \eqref{M2*brane1} to map it into an $M$-theory solution. However the $M2$ root, $\vec{\alpha}_{11}$, is invariant under this Weyl reflection as $<\vec{\alpha}_{11},\vec{\beta}_{129}>=0$. Consequently the Weyl reflection has a trivial action on the group element encoding the $M2^*$ solution but it does change the background signature of space-time. While we have observed that the background signature is modified by the Weyl reflection's action on the signature function, it will be useful to emphasise this in more detail. The isometries of space-time are encoded in the level zero involution invariant sub-algebra of $K(E_{11})$ whose generators are $Q^0_i\equiv {K^i}_{i+1}-\epsilon_i {K^{i+1}}_{i}$, where $\epsilon_i$ are defined in equation (\ref{involution}). When one of $x^i$ and $x^{i+1}$ is temporal and the other is spatial $\exp{(\theta Q^0_i)}$ is a (non-compact) boost parameterised by $\theta$, while, if both coordinates are temporal or both are spatial, the corresponding group element is a (compact) rotation. Under a Weyl reflection the generators of $E_{11}$ may be interchanged and the properties of the $Q^0_i$ generators may be changed; specifically, in the local group, if a boost is mapped to a rotation or vice-versa then there is a change in the signature of the space-time. Under the Weyl reflection $S_{\beta_{129}}$ the $Q^0_i$ are unchanged apart from:
\begin{align}
Q^0_2={K^2}_{3}+ {K^{3}}_{2}	 &\to R_{139}+ R^{139} 	\\
Q^0_8={K^8}_{9}- {K^{9}}_{8}	 &\to R^{128}- R_{128} 	\\
Q^0_9={K^9}_{10}- {K^{10}}_{9}	 &\to R_{1210}- R^{1210}.
\end{align}
The generators of boosts and rotations in space-time are mapped to elements of ${\cal K}(E_{11})$ appearing at level one, and, while the space-time signature is changed, the compact or non-compact nature of each algebraic element is unchanged by the Weyl reflection. The level one local transformation $Q^1_k$ acts on the 55 coordinates $y_{ab}$ at level one in the $l_1$ representation of $E_{11}$. The extra coordinates have been interchanged with the usual space-time coordinates as
\begin{align}
x^1&\longleftrightarrow y_{29}	\\
x^2&\longleftrightarrow y_{19}	\\
x^9&\longleftrightarrow y_{12}.	
\end{align}
The wavefunctions of the $M2^*$ solution depend on $\vec{k}\cdot\vec{x}+\vec{\omega}\cdot\vec{t}$ where $\vec{t}=(x^1,x^2)^T$ and $\vec{x}=(x^3,x^4,\ldots x^8)^T$; under the reflection $S_{\beta_{129}}$, $\vec{t}\to (y_{29},y_{19})^T$ and $\vec{x}\to\vec{x}$. Note that, following the change in signature, both $y_{29}$ and $y_{19}$ are timelike coordinates. {\cred Hence the $M2^*$-brane is mapped to an $M$-theory solution in an extension of supergravity which depends explicitly upon extra coordinates of the form $y_{\mu_1\mu_2}$ and is defined by wavefunctions rather than harmonic functions.} The part of the metric in the usual eleven-dimensional space-time is:
\begin{align}
ds^2=&(f+g)^{1/3}((dx^1)^2+(dx^2)^2+(dx^3)^2\ldots +(dx^8)^2)+ \nonumber\\
& (f+g)^{-2/3}(-(dx^9)^2+(dx^{10})^2+(dx^{11})^2) \label{GeneralisedM2brane1}
\end{align}
where now $f(\vec{x}\cdot\vec{k}+\omega\cdot \vec{t})$ and $g(\vec{x}\cdot\vec{k}-\omega\cdot \vec{t})$, where $\vec{\omega}=(\omega_1,\omega_2)^T$, $\vec{k}=(k_1,k_2,\ldots, k_6)^T$, $\vec{t}=(y_{29},y_{19})^T$ and $\vec{x}=(x^3,x^4,\ldots, x^8)^T$. The dispersion relation is $\vec{\omega}^2={\vec{k}}^2$ and the non-trivial components of the field strength are
\begin{equation}
F_{\hat{\Sigma} \hat{9}\hat{10}\hat{11}}=\partial_{\hat{\Sigma}} (\frac{-1}{f+g}) \label{GeneralisedM2brane2}
\end{equation}
where $\partial_{\hat\Sigma}$ denotes derivatives with respect to the coordinates $\{y_{29},y_{19}\}$ and $\{x^3,x^4,x^5,x^6,x^7,x^8\}$.
\subsection{Explicit check of the eleven-dimensional field equations.}
\label{sec:EOMcheck}

{ \cred In this subsection we recall the direct space-time check that the
travelling-wave membrane-type solutions generated by the two-parameter
$\sigma$-model satisfy the bosonic field equations in the corresponding
eleven-dimensional signature. This provides a useful consistency check of
the embedding, in addition to the coset construction.}

We consider the standard bosonic equations (with signature-dependent signs
suppressed for clarity):
\begin{align}
R_{MN} &= \frac{1}{12}\left(F_{MPQR}{F_N}^{PQR}
-\frac{1}{12}g_{MN}F_{PQRS}F^{PQRS}\right),
\label{eq:Einstein11}\\
dF &= 0, \label{eq:Bianchi}\\
d\star F &= 0. \label{eq:Maxwell11}
\end{align}

\paragraph{Ansatz.}
We take the warped metric and four-form field strength appropriate to the
membrane sector,
\begin{align}
ds^2 &= N^{1/3} ds^2_{\perp} + N^{-2/3} ds^2_{\parallel}, \label{eq:metric_ansatz}\\
F &= d(N^{-1}) \wedge \mathrm{vol}_{\parallel}, \label{eq:F_ansatz}
\end{align}
where $ds^2_{\parallel}$ is the induced metric on the three world-volume
directions and $ds^2_{\perp}$ is the metric on the transverse space.
In contrast to the standard static brane solutions, the warp factor
\begin{equation}
N = f(\sigma+\tau)+g(\sigma-\tau)
\end{equation}
depends on two transverse coordinates of mixed signature.

\paragraph{Bianchi identity.}
Since $F$ is exact by construction, the Bianchi identity \eqref{eq:Bianchi}
is automatically satisfied.

\paragraph{Maxwell equation.}
The Maxwell equation \eqref{eq:Maxwell11} reduces to
\begin{equation}
d\star\left[d(N^{-1}) \wedge \mathrm{vol}_{\parallel}\right]=0.
\end{equation}
Using the warped product structure \eqref{eq:metric_ansatz}, this equation
reduces to
\begin{equation}
\Box_{\perp} N = 0,
\label{eq:waveN}
\end{equation}
where $\Box_{\perp}$ is the Laplace--Beltrami operator associated with
$ds^2_{\perp}$. In a transverse space with Lorentzian signature in the
$(\tau,\sigma)$ plane, \eqref{eq:waveN} is a wave equation rather than a
Laplace equation. The travelling--wave form of $N$ therefore solves the
Maxwell equation identically.

\paragraph{Einstein equation.}
For warped ansätze of the form \eqref{eq:metric_ansatz}, the Einstein
equations \eqref{eq:Einstein11} reduce to algebraic relations between warp
exponents together with the same differential condition \eqref{eq:waveN}
on $N$. The relative powers $N^{1/3}$ and $N^{-2/3}$ are precisely those
required for cancellation between curvature terms and stress-energy terms
constructed from \eqref{eq:F_ansatz}. Consequently, the Einstein equations
are satisfied provided \eqref{eq:waveN} holds.
\paragraph{Conclusion.}
The two-parameter $\sigma$-model constraints enforce exactly the same wave
equation on $N$ that arises from the eleven-dimensional field equations.
{\cred Thus, for the membrane-type ansatz considered here, the travelling-wave
condition obtained from the two-parameter $\sigma$-model is precisely the
condition required by the classical bosonic field equations in the
appropriate signature.}

\cb{\subsection{Gravitational interpretation of the two-parameter solutions}
\label{subsec:grav_interpretation_twoparam}

Let us finally reconsider the gravitational meaning of the two-parameter
solutions constructed in this paper. As noted earlier, the two-dimensional sigma-model used
above is not intended to define a fundamental worldsheet quantum field
theory.  Its purpose is instead analogous to that of the one-parameter brane
sigma-model: it is an auxiliary, solution-generating device.  Once the
sigma-model fields are embedded into spacetime, the resulting configurations
are to be tested as classical solutions of the appropriate bosonic field
equations.

In the one-parameter brane sigma-model, the scalar function appearing in the
coset solution becomes the familiar harmonic function of the corresponding
brane background.  In the present two-parameter extension the corresponding
function is instead a solution of a wave equation.  {\cred For the genuinely two-parameter solution found in case (v) of
Section~\ref{case5}, one has}
\begin{equation}
  N(\sigma,\tau)
  =
  f(\sigma+\tau)+g(\sigma-\tau),
  \label{eq:N_travelling_wave_conclusion}
\end{equation}
where $f$ and $g$ are arbitrary left- and right-moving functions.  After
embedding, this function plays the role of the brane warp factor.  For
example, in the membrane-type case one obtains a metric of the schematic
form
\begin{equation}
  ds^2
  =
  N^{1/3} ds^2_{\perp}
  +
  N^{-2/3} ds^2_{\parallel},
  \label{eq:brane_metric_schematic_conclusion}
\end{equation}
together with a four-form field strength of the form
\begin{equation}
  F
  =
  d\!\left(N^{-1}\right)\wedge \mathrm{vol}_{\parallel}.
  \label{eq:brane_flux_schematic_conclusion}
\end{equation}
The field equations then reduce to the condition that $N$ solve the
appropriate transverse equation.  In the ordinary single-time brane
solutions this is a Laplace equation,
\begin{equation}
  \nabla^2_{\perp} N = 0,
  \label{eq:laplace_standard_conclusion}
\end{equation}
and hence $N$ is a harmonic function.  By contrast, in the present
two-parameter construction the relevant transverse sector contains one
spacelike and one timelike coordinate.  The corresponding equation is
therefore
\begin{equation}
  \Box_{\perp} N
  =
  \left(
    \partial^2_{\sigma}
    -
    \partial^2_{\tau}
  \right)N
  =
  0,
  \label{eq:wave_equation_conclusion}
\end{equation}
which is solved precisely by the travelling-wave expression
\eqref{eq:N_travelling_wave_conclusion}.

This observation gives the two-parameter construction its direct
gravitational interpretation.  The solutions are not arbitrary functions
placed by hand into a metric ansatz.  Rather, they arise from the
sigma-model equations and, after embedding, solve the classical bosonic
field equations in the appropriate signature sector.  The replacement
\begin{equation}
  \hbox{harmonic function}
  \quad \longrightarrow \quad
  \hbox{travelling wavefunction}
  \label{eq:harmonic_to_wave_conclusion}
\end{equation}
is therefore the spacetime manifestation of replacing the one-parameter
null geodesic on the symmetric space by a two-parameter worldsheet on the
same symmetric space.

This also explains why the natural spacetime interpretation of these
solutions involves $M^\ast$- and $M'$-theory.  In ordinary M-theory the
transverse space of the standard electric branes is Riemannian, and the
brane function is harmonic.  The two-parameter solutions instead require a
Lorentzian transverse two-plane with an $SO(1,1)$ invariant metric.  Such
a sector is naturally available in the multiple-time signatures associated
with $M^\ast$- and $M'$-theory.  Thus the appearance of these theories is
not an additional assumption, but follows from the requirement that the
two-parameter coset solution be embedded as a classical gravitational
background.

Finally, the $E_{11}$ interpretation is sharpened by applying Weyl
reflections.  The wavefunction-dependent solutions may be written naturally
in the $M^\ast$- or $M'$-theory frames using ordinary spacetime
coordinates adapted to the multiple-time signature.  After an $E_{11}$
Weyl reflection back to the ordinary M-theory frame, the same solution is
mapped to one whose natural coordinate dependence is no longer purely on
the conventional eleven coordinates $x^\mu$.  Instead, the dependence is
transferred to the extended coordinates belonging to the fundamental
$l_1$ representation of $E_{11}$, such as
\begin{equation}
  y_{\mu_1\mu_2},
  \qquad
  z_{\mu_1\cdots\mu_5},
  \qquad
  w_{\mu_1\cdots\mu_7|\nu},
  \qquad
  \ldots .
  \label{eq:extended_coordinates_conclusion}
\end{equation}
In this sense, the two-parameter solutions provide an explicit
solution-level link between three ideas: travelling waves on symmetric
spaces, classical gravitational backgrounds in $M^\ast$- and
$M'$-theory, and extended-coordinate dependence in the $E_{11}$
description of M-theory.  The construction therefore supports the view that
the extended coordinates are not merely formal algebraic artefacts, but are
required for the M-theory interpretation of certain Weyl-reflected
classical solutions.

\cbl

\subsection{Classification of two-parameter solutions.}
\label{sec:classification}

For clarity, we summarise the classes of two-parameter solutions discussed
above and their physical interpretation in table \ref{tab:classification}. The table highlights that the principal new physical content of the
two-parameter extension resides in case (v), where the harmonic function
of standard brane solutions is replaced by a travelling wave. All other
cases either reduce to the one-parameter geodesic solutions or fail to
admit a consistent higher-dimensional embedding.

\begin{table}[th]
\centering
\small   % ← optional: makes the font a bit smaller → helps a lot with fitting
\begin{tabularx}{\linewidth}{|>{\centering\arraybackslash}p{1.0cm}|>{\raggedright\arraybackslash}p{3cm}|>{\centering\arraybackslash}p{2.2cm}|>{\centering\arraybackslash}p{2.2cm}|>{\centering\arraybackslash}X|}
\hline
Case & $\phi(\tau,\sigma)$, $C(\tau,\sigma)$ & Genuine 2D? & Embedding & Warp factor \\ 
\hline
(i)   & $\phi(\sigma)$, $C(\sigma)$ 
      & No 
      & $M^*$, $M'$ 
      & Linear reductions \\ 
\hline
(ii)  & $\phi(\tau)$, $C(\tau)$ 
      & No 
      & $M^*$, $M'$ 
      & Linear reductions \\ 
\hline
(iii) & Mixed but constrained
      & No 
      & Obstructed 
      & --- \\ 
\hline
(iv)  & Special 2D-only solutions
      & Yes (2D only) 
      & No higher-$D$ lift 
      & --- \\ 
\hline
(v)   & $\displaystyle\phi = \tfrac{1}{2}\ln(f+g)$, $C = -(f+g)^{-1}$
      & Yes 
      & $M^*$, $M'$ 
      & Travelling wave \\ 
\hline
\end{tabularx}

\caption{Classification of two-parameter $\sigma$-model solution types. 
Only case (v) gives a genuinely two-parameter higher-dimensional brane
background with a wave-like warp factor.}
\label{tab:classification}
\end{table}

\section{Conclusions}\label{conclusions}
In this paper we have explored a minimal extension of the brane $\sigma$-model on $\frac{SL(2,\mathbb{R})}{SO(1,1)}$ which replaces the point particle motion with a string motion on the symmetric space. The extension allows the geometry of the two-dimensional symmetric space to be probed. We took a conservative approach to the model extension: we worked in the flat metric in the coset, and retained all features of the original point particle brane $\sigma$-model. We constructed solutions dependent on two parameters, one of which is temporal and the other spatial, and demonstrated the model gave novel solutions to $M^*$ and $M'$-theories.

The organising principle of the paper is to work with a minimal extension of the brane $\sigma$-model as {\em a solution-generating framework}, rather than as a proposal for a fundamental two-dimensional field theory. In particular, its role is to probe the structure of the symmetric space underlying known supergravity solutions. The original, one-parameter brane $\sigma$-model derived brane solutions of $M$-theory from the root system of $E_{11}$: the brane solutions in space-time had an alternative description as the null geodesic worldline of a particle on the symmetric space. Solving the equations of the model gives harmonic functions dependent on a single parameter, which can then be embedded in the M-theory spacetime with the parameter being identified with a spatial coordinate transverse to the world-volume of a brane solution. In the extension presented here solutions of $M^*$-theory and $M'$-theory are derived from open string worldsheets on the symmetric space. These worldsheets should be understood as geometrical probes of the coset structure rather than as dynamical strings propagating in a fixed target space. The equations of motion of the model are solved by wavefunctions, one for left-moving waves and the other for right-moving waves. The wavefunctions depend on two parameters, one spacelike and one timelike. Embedding such solutions into spacetime therefore requires both a spacelike and a timelike transverse coordinate. The brane world-volume already includes a temporal coordinate, thus the embedding is possible in theories which contain multiple time coordinates, such as the $M^*$ and $M'$-theories which are related to the $M$-theory sector of $E_{11}$ by Weyl reflections.

By following the action of a suitable Weyl reflection on the $M2^*$ brane, we argued that the corresponding $M$-theory solution depends not only on the usual space-time coordinates $x^\mu$, but also on the $y_{\mu\nu}$ coordinates which arise in the $l_1$ representation of $E_{11}$. It is expected that the solutions presented here in terms of wavefunctions are admitted within $M$-theory once the extended coordinate system of $E_{11}$ is used, for among the $y_{\mu\nu}$ coordinates are ten which are timelike and can play the role of the extra temporal coordinates in $M^*$ and $M'$-theory. To test the conjecture that $M$-theory solutions dependent upon extra coordinates are dual to the $M^*$ and $M'$-theory solutions requires an extension of the supergravity action to include space-time constructed out of eleven $x^\mu$ and fifty-five $y_{\mu\nu}$ coordinates. Such a theory would be sufficient to study sectors of $M$, $M^*$ and $M'$-theories in a single setting carrying the low-level symmetries of $\frac{E_{11}}{{\cal K}(E_{11})}$. The solutions presented in this paper offer a guide for constructing the enlarged theory. 

To make these ideas more concrete, we briefly outline the symmetry structure that would arise in such an extension. The extension of supergravity to incorporate $y_{\mu\nu}$ leads to a sixty-six-dimensional theory. The temporal and spatial interpretation of the $y_{\mu\nu}$ coordinates are derived from the properties of the coordinates $x^\mu$, hence for a background space-time with an $SO(1,10)$ isometry on the $x^\mu$ coordinates the $y_{\mu\nu}$ background has an $SO(10,45)$ isometry. There are no rotations of $x^\mu$ coordinates into $y_{\mu\nu}$ coordinates, so the isometries of the background space-time are $SO(1,10)\times SO(10,45)$. If we assume that the theory is translation invariant in both $x^\mu$ and $y_{\mu\nu}$ and carries the Lorentz symmetries $SO(1,10)$ and $SO(10,45)$ in the two sets of coordinates then we may make some initial observations about the structure of the theory. We adopt the commutators suggested by the canonical embedding of the $l_1$ representation into the algebra of $E_{12}$, namely,
\begin{align}
[P_a, {K^b}_c]&=\delta^b_a P_c-\frac{1}{2}\delta^b{}_cP_a,\\	
[P_a,R^{bcd}]&=3\delta_a^{[b}Z^{cd]},\\
[{K^a}_b,Z^{cd}]&= 2\delta_b^{[c}Z^{|a|d]}-\frac{1}{2}\delta^a{}_bZ^{cd}\\
[P_a,P_b]&=0, \quad [Z^{ab},Z^{cd}]=0 \quad \mbox{ and } \quad [P_a,Z^{bc}]=0.
\end{align}
Under translations and rotations on the coordinates $(x^\mu,y_{\mu\nu})$ are mapped to $(x'^\mu,y'_{\mu\nu})$ according to
\begin{align}
&\exp(a^\mu P_\mu + {u_\mu}^\nu {(Q_0)^\mu}_\nu + b_{\mu\nu}Z^{\mu\nu}+v_{\mu\nu\rho}(Q_1)^{\mu\nu\rho})\exp(x^\mu P_\mu + y_{\mu\nu}Z^{\mu\nu}) \nonumber \\
&=\exp(x'^\mu P_\mu + y'_{\mu\nu}Z^{\mu\nu})\exp({u'_\mu}^\nu {(Q_0)^\mu}_\nu +v'_{\mu\nu\rho}(Q_1)^{\mu\nu\rho})
\end{align}
where ${(Q_0)^\mu}_\nu\equiv {K^\mu}_\nu-\Omega({K^\mu}_\nu)$ and $(Q_1)^{\mu\nu\rho}\equiv R^{\mu\nu\rho}-\Omega(R^{\mu\nu\rho})$ and $\Omega$ is the temporal involution defined on $E_{11}$. Hence, for infinitesimal transformations, we find the action must be invariant under
\begin{align}
\delta x^\mu &= a^\mu -{u_\nu}^\mu x^\nu	\\
\delta y_{\mu\nu} &= b_{\mu\nu} +2{u_\mu}^\kappa y_{\kappa\nu}-3x^\kappa v_{\kappa\mu\nu}.
\end{align}
The final term in $\delta y_{\mu\nu}$ is novel; the other terms arise from the $SO(1,10)$ Lorentz transformations and translations. We note that this transformation (associated with the $SO(10,45)$ Lorentz transformation) leads to a variation of the Lagrangian density for kinetic terms, i.e. under $\delta y_{\mu\nu} = x^\kappa v_{\kappa\mu\nu}$ the Lagrangian density
\begin{equation}
{\cal L}=\frac{1}{2}\partial_\mu \phi \partial^\mu \phi +\frac{1}{2}\partial_{\mu\nu} \phi \partial^{\mu\nu} \phi 
\end{equation}
has a non-vanishing variation
\begin{equation}
\delta {\cal L}=v_{\kappa\mu\nu}\partial^{\kappa\nu}\phi \partial^\mu \phi
\end{equation}
upto total derivative terms. In two-dimensions, as on the coset manifolds we have looked at in this paper, the shift parameter $v_{\kappa\mu\nu}$ is zero as it is an antisymmetric tensor. However when working on larger symmetric spaces, such as $\frac{SL(3)}{SO(2,1)}$ which is used to reconstruct bound states of supergravity branes \cite{Cook:2009ri,Houart:2009ya} and whose dimension is greater than three, corrections to the $\sigma$-model Lagrangian will be needed to ensure invariance under translations in $y_{\mu\nu}$. Generalising the $\sigma$-model construction so that it depends on more than two parameters, and can be used to construct solutions dependent on the extended coordinates of the $l_1$ representation of $E_{11}$, is an interesting direction for future work.

Our work argues in favour of an extension to $M$-theory to a theory which includes $M^*$-theory and $M'$-theory as well and set in a space-time constructed from the coordinates derived from the $l_1$ representation of $E_{11}$. A necessary consequence of our observations is that solutions in M-theory must exist which depend non-trivially on the ``exotic coordinates" $y_{\mu\nu}$, $z_{\mu\nu\rho\sigma\tau}$ and so on. Recently there has been significant progress in constructing solutions which depend on extra coordinates in the settings of double field theory (DFT) and exceptional field theory (EFT), large parts of each construction may be understood as being derivable from the $E_{11}$ framework for M-theory. A large class of solutions to these theories have been investigated in \cite{Berkeley:2014nza,Berman:2014hna,Berman:2014jsa,Bakhmatov:2016kfn,Blair:2016xnn} where solutions have been constructed which depend on an extra coordinate. The search for solutions to DFT and EFT benefits from the `section condition' which projects out extra coordinates dependent upon the choice of duality frame. There remains some work to do to relate the construction of solutions from a brane $\sigma$-model to the solutions found from DFT and EFT. Two issues appear particularly immediate in this respect. Firstly, while the idea of the duality frame can be interpreted naturally as singling out the sub-algebra of $E_{11}$ used to define the symmetric space on which the $\sigma$-model is constructed, there has yet to be any work done on investigating the brane $\sigma$-model when the coordinates of the $\sigma$-model are identified with exotic coordinates. Secondly there is no necessary requirement for the section-condition from the $E_{11}$ point of view, while in DFT and EFT it plays a crucial role in simplifying the dependence of the theory on the extra coordinates to the point that solutions to the equations of motion may be constructed \cite{Berkeley:2014nza,Berman:2014hna,Berman:2014jsa}. It seems sensible for future work that the section-condition is given an interpretation within the brane $\sigma$-model setting as a consistency check on the two approaches to the infinite-dimensional space-time of $M$-theory\footnote{See the recent paper \cite{Bossard:2017wxl} where the section condition has been adopted within the $E_{11}$ framework.}. 

The causality of solutions in theories with multiple time coordinates has not been addressed in this work. However, the solutions of the $M^*$ and $M'$-theories that we have constructed have all been mapped, via Weyl reflections, to sectors of $M$-theory, where questions of causality are minimised. Solutions of $M$-theory for which the evolution of a Cauchy surface is well-defined, and hence possess a determinate causal structure, are mapped by Weyl reflections to solutions of $M^*$ and $M'$ theory which depend on exotic coordinates\footnote{The reverse direction of the mapping of solutions has been focussed on in the present paper, specifically, solutions in $M*$ and $M'$ theories dependent on $x^\mu$ co-ordinates were constructed and mapped by Weyl reflections to solutions of $M$-theory dependent on a mixture of eleven of the $x^\mu$ and $y_{\mu\nu}$ coordinates. However the argument made here is a consequence of these observations.}. The maps are bijections and no information is lost in the mapping solutions between theories with different signatures. Consequently it is expected that the causal structure of $M$-theory may be preserved in the map of its solutions into $M^*$ and $M'$ theories. It would be an interesting future work to investigate the causal structure of $M^*$ and $M'$ theories and one may begin the work by using $E_{11}$ to map identify the equivalent of the Cauchy surfaces in the theories with multiple time coordinates.

\cb{The solutions constructed here are classical solutions of the bosonic
field equations obtained after embedding the sigma-model data into the
corresponding spacetime fields.  This is the same sense in which
ordinary brane backgrounds are classical solutions relevant to string
theory or M-theory: they describe semiclassical backgrounds of the
low-energy effective theory, not the full quantum theory.  The
high-energy significance of the solutions lies in their transformation
properties under $E_{11}$.  Since $M$, $M^\ast$ and $M'$ arise as
different signature sectors related by Weyl reflections, the classical
solutions in the multiple-time frames are not independent ad hoc
backgrounds, but representatives of $E_{11}$-related solution
classes.  Their Weyl-reflected images in the M-theory frame require
dependence on extended coordinates, thereby exposing a sector of the
classical solution space which is invisible in the usual
eleven-coordinate truncation.}\cbl

The work in this paper probed an extension of the standard (one parameter) brane $\sigma$-model construction. The approach to the brane $\sigma$-model, described in the introduction to this paper, tempted the authors to explore whether one could generalise the approach to find two-parameter solutions. We took a minimal approach to the extension, in particular although we were considering string world sheets on two-dimensional symmetric spaces, we opted to work with the flat Minkowski metric, valid only in a small neighbourhood of the identity in the computations in section 3. The symmetric space $\frac{SL(2,\mathbb{R})}{SO(1,1)}$ has a non-trivial topology, being diffeomorphic to the one-sheeted hyperboloid. The benefit of the using the flat metric was the simplicity of the equations emerging from the $\sigma$-model, but it was at a cost of not exploring the role of the metric, and the topology, of the symmetric space. The restriction to the flat metric is not the only bar to exploring the topology of the symmetric space on the putative space-time solutions. In appendix \ref{Borelgauge} we have described the effect that working in the Borel gauge has on eliminating the topology of $SL(2,\mathbb{R})/SO(1,1)$ in the construction of solutions. The abstract string world-sheet found by solving the equations of motion of (\ref{so11sigmamodel}) traverses a subspace of the symmetric space which is isomorphic to $\mathbb{R}^{1,1}$. The full $S^1\times \mathbb{R}$ topology of $SL(2,\mathbb{R})/SO(1,1)$ is lost due to working with the Borel gauge. It is therefore an interesting direction of research to consider alternative gauge choices, such as the unitary gauge, in conjunction with the full metric of the symmetric space that will enable the topology of the symmetric spaces to play a role: it would be particularly interesting to consider closed string world-sheets moving on $SL(2,\mathbb{R})/SO(1,1)$ and wrapping the $S^1$.

There remains a great deal of work to be done in investigating the role of exotic coordinates in physical theories. A central challenge will be the construction and interpretation of a broader class of brane solutions dependent on the enlarged, infinite-dimensional space-time.

\section*{Acknowledgements}
The authors are grateful to Michael Duff and David Berman for their remarks on this work. PPC would like to thank the Department of Mathematics at the University of Bath for their hospitality while part of this work was carried out. We wish to thank the STFC for their support under the consolidated grant number ST/X000753/1. SS would like to thank Philip Mannheim for a helpful discussion.
{\cred
\section*{Data availability statement}
No new data were generated or analysed in this study. 
}
\appendix
\section{The Borel Gauge and $\frac{SL(2,\mathbb{R})}{SO(1,1)}$} \label{Borelgauge}
In this appendix we describe the topology of the coset space $\frac{SL(2,\mathbb{R})}{SO(1,1)}$ and illustrate the paths traversed by the one-parameter brane solutions with a focus on how constrained these paths are by the use of the Borel gauge.

First consider a matrix $M\in SL(2,\mathbb{R})$ given by
\begin{equation}
M\equiv\exp(aH+b(E-F))=
\left(
\begin{array}{cc} 
\cosh(r)+\frac{a}{r}\sinh(r)& \frac{b}{r}\sinh(r) \\ 
-\frac{b}{r}\sinh(r) & \cosh(r)-\frac{a}{r}\sinh(r)
\end{array}
\right)
\end{equation}
where $r^2=a^2-b^2$. The elements of the coset $\frac{SL(2,\mathbb{R})}{SO(1,1)}$ are given by 
$$g'=M \exp(c(E+F)).$$
Now writing $x=\frac{b}{r}\sinh(r)$, $y=\frac{a}{r}\sinh(r)$ and $z=\cosh(r)$ then from $\det(M)=1$ we have $x^2-y^2+z^2=1$ which is a single-sheeted hyperboloid.

Our aim is to illustrate the null geodesic which encodes the brane solution on the representative hyperboloid for the coset. Recall that each brane solution is given by a representative group element for the coset in the Borel (upper triangular) gauge, explicitly,
\begin{align*}g&=\exp\left(
\begin{array}{cc} 
\frac{1}{2}\ln(N) & 0 \\ 
0 & -\frac{1}{2}\ln(N)
\end{array}\right)\exp\left(
\begin{array}{cc} 
0 & 1-N^{-1} \\ 
0 & 0
\end{array}\right)\\
&=\sqrt{N}\left(
\begin{array}{cc} 
1 & 1-N^{-1} \\ 
0 & N^{-1}
\end{array}\right)
\end{align*}
where $N=1+q\xi$. In the above presentation of the group element one of the constants of integration in the generic solution has been chosen such that when $\xi=0$ $g=\mathbb{I}$, the identity element - which upon embedding the group element into space-time corresponds to Minkowski space. 

The matrix $M$ above parameterises the hyperboloid $x^2-y^2+z^2=1$ and compared with the brane solution group element $g$ it is written in a different gauge. The action of $SO(1,1)$ in the coset may allow $M$ to be written in the Borel gauge, this amounts to choosing $c$ above such that $g'$ is an upper triangular matrix, 
\begin{align*}
g'&=\left(
\begin{array}{cc} 
z+y& x \\ 
-x & z-y
\end{array}
\right)\left(
\begin{array}{cc} 
\cosh(c) & \sinh(c) \\ 
\sinh(c) & \cosh(c)
\end{array}
\right)\\
&=\left(
\begin{array}{cc} 
(z+y)\cosh(c)+x\sinh(c)& (z+y)\sinh(c)+x\cosh(c) \\ 
(z-y)\sinh(c)-x\cosh(c) & (z-y)\cosh(c)-x\sinh(c)
\end{array}
\right)
\end{align*}
i.e. to put $g'$ in Borel gauge requires choosing $c$ such that
$$(z-y)\sinh(c)-x\cosh(c)=0$$
that is,
$$\tanh(c)=\frac{x}{z-y}.$$ 
Hence the coordinates ${x,y,z}$ on the hyperboloid $x^2-y^2+z^2=1$ are constrained such that $\frac{x}{z-y}\in (-1,1)$ when the representative group element is in the Borel gauge. For example, in the plane $y=0$ the cross-section of the hyperboloid is given by $x^2+z^2=1$ and the coordinates are constrained by the Borel gauge such that $-z<x<z$. This gives two disconnected line elements, one of which includes the identity element (the point $x=0$, $y=0$, $z=1$). In general for non-zero $y=y_0$, the cross-section of the hyperboloid is the circle $x^2+z^2=1+y_0^2$ and the choice of Borel gauge constrains the $x$ and $z$ coordinates to satisfy $-(z-y_0)<x<(z-y_0)$. The Borel gauge constrains the group elements to two disconnected set of points each topologically equivalent to $\mathbb{R}^2$, only one of which is connected to the identity element. As observed in \cite{Keurentjes:2005jw}, the use of the Borel gauge means that the topology of $\frac{SL(2,\mathbb{R})}{SO(1,1)}$ is reduced to $\mathbb{R}^2$. The loss of information from closed cycles in cosets embedded in $E_{11}$ motivates considering the two-parameter coset model described in the present paper.

We now identify the path of the brane solution, as $\xi$ varies, on the hyperboloid $x^2-y^2+z^2=1$. To do this we first write $g'$ in Borel gauge by substituting $\cosh(c)=\frac{\pm(z-y)}{\sqrt{2z(z-y)-1}}$ and $\sinh(c)=\frac{\pm x}{\sqrt{2z(z-y)-1}}$ \footnote{The signs may be fixed for given $y$ and $z$ coordinates by the positivity of $\cosh(c)$.} to obtain
$$g'=\frac{\pm 1}{\sqrt{2z(z-y)-1}}\left(
\begin{array}{cc} 
1 & 2xz\\ 
0 & 2z(z-y)-1
\end{array}
\right).$$
By comparing this matrix with that for the solution encoding group element $g$ we find 
$$N^{-1}=2z(z-y)-1 \qquad \mbox{and} \qquad 2xz=1-N^{-1}.$$
Hence the solution is given by the intersection of 
$$x-y=\frac{1-z^2}{z}$$
and the hyperboloid $x^2-y^2+z^2=1$. The intersection points satisfy
$$(z-1)(z+1)(2z^2-2yz-1)=0$$
and only the solution where $z=1$ for which $x=y$ passes through the identity element and this corresponds to the brane solution. We note that the line of points on the hyperboloid such that $2z^2-2yz-1=0$ corresponds to a constant gauge field (as $z=\frac{1}{2x}$) but gives an infinite value to the harmonic function $N$. We have illustrated the lines of intersection on the hyperboloid in figure \ref{linesonhyperboloid}.

\begin{figure}[h]
\centering\includegraphics[scale=0.5]{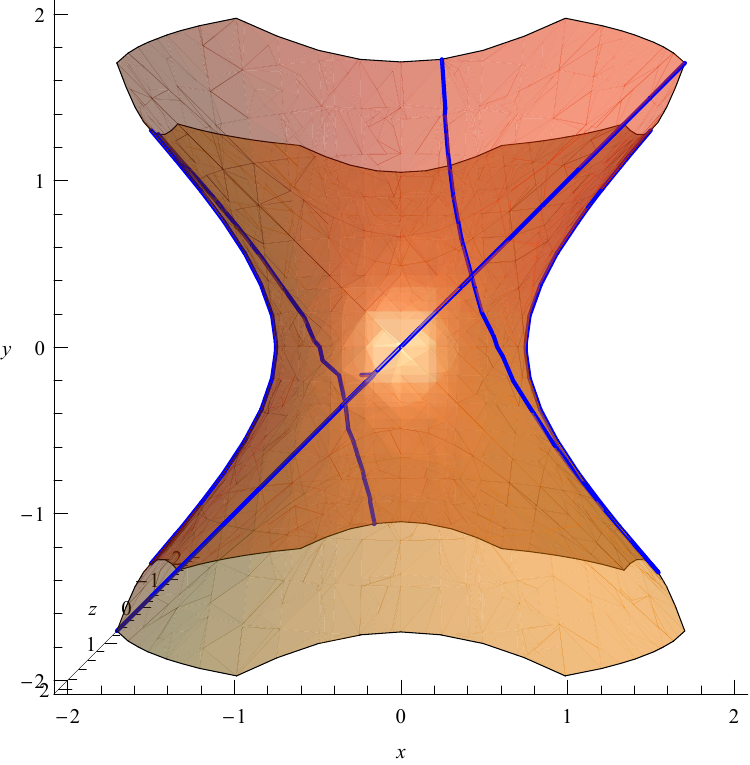}
\caption{The lines of intersection of $x^2-y^2+z^2=1$ and $x-y=\frac{1-z^2}{z}$: $x=y$ and $z=-1$ or $z=1$, and $2z^2-2yz=1$. The brane solution corresponds to points on the line $x=y$ in the plane $z=1$, and the point $(0,0,1)$ corresponds to Minkowski space.} \label{linesonhyperboloid}
\end{figure}

Returning to the brane solution, given by the line of points $x=y$ in the plane $z=1$, we note that from $N=1+q\xi$ we may read off $\xi(x,y,z)$ as $N^{-1}=2z(z-y)-1=1-2y$, hence $q\xi =\frac{2x}{1-2x}$. The map between coordinates is ill-defined at $x=\frac{1}{2}$ and as $x\to\pm \infty$ then $q\xi\to -1$. The range $x\in[0,\frac{1}{2})$ corresponds to $\xi\in [0,\infty]$, with $\xi=0$ being Minkowski space. In summary the brane solution corresponds to the points $(x,x,1)$ where $x\in [0,\frac{1}{2})$, which lie on the hyperboloid $x^2-y^2+z^2=1$ representing the coset $\frac{SL(2,\mathbb{R})}{SO(1,1)}$.In other words, of the full hyperboloid only a short line segment plays any role in defining the brane solution in the one-parameter brane $\sigma$-model and the topology of the symmetric space plays no role in the solution embedded in space-time. These observations underline and motivate the need to explore extensions to the one parameter $\sigma$-model.

\section{The reduction of the $4-$dimensional Einstein action to an action in terms of the $3$-metric} \label{metricreduction}

In $4$-dimensional space-time the Einstein-Hilbert action can be written as
 \begin{equation}
\label{EH }
\int d^{4}x \sqrt{-g}L\left(g_{\mu\nu},\partial_{\gamma}g_{\mu\nu}\right)
\end{equation}
where $L$ is written as 
\begin{equation}
\label{EH2 }
L=R
\end{equation}
where $R$ is the Ricci scalar (and we have excluded any cosmological constant and constant of proportionality). Moreover 
\begin{equation}
\label{ ricci}
R=q_{\mu}\,^{\nu\lambda\delta}R^{\mu}\,_{\nu\lambda\delta}
\end{equation}
with $q_{\mu}\,^{\nu\lambda\delta}=\frac{1}{2}(\delta_{\mu}^{\lambda}g^{\nu\delta}-\delta_{\mu}^{\delta}g^{\nu\lambda})$. We note that 
\begin{equation}
\sqrt{-g}L=2\partial_{\gamma}\left[\sqrt{-g}q_{\alpha}\,^{\beta\gamma\delta}\Gamma^{\alpha}_{\beta\delta}\right]+2\sqrt{-g}q_{\alpha}^{\beta\gamma\delta}\Gamma^{\alpha}_{\delta\kappa}\Gamma^{\kappa}_{\beta\gamma}.\label{surface}
\end{equation}
We shall ignore the total derivative term in (\ref{surface}) and so the new form of the the Einstein-Hilbert action (which does not contain any terms which contain second derivatives in the metric) is
\begin{equation}
\label{EH3 }
-\int d^{4}x\sqrt{-g}\mathcal{G}
\end{equation}
where 
\begin{equation}
\label{EH4 }
\mathcal{G}=g^{\mu\nu}\left( \Gamma^{\alpha}_{\mu\beta}\Gamma^{\beta}_{\nu\alpha}-\Gamma^{\alpha}_{\mu\nu}\Gamma^{\beta}_{\alpha\beta}\right).
\end{equation}

In order to rewrite $\mathcal{G}$ in terms of metrics on space-like foliations of space-time it is convenient to work in Gaussian co-ordinates. Any space-like hypersurface $S$ in this foliation will be intersected orthogonally by a family of geodesics. The length along these geodesics will give the time. It is then consistent to write the metric in terms of a $3$-metric $\gamma_{ij}$ ($1\leqslant i,j \leqslant 3$) as follows:

\[
\left(\begin{array}{cccc}
1 & 0 & 0 & 0\\
0 & \gamma_{11} & \gamma_{12} & \gamma_{13}\\
0 & \gamma_{21} & \gamma_{22} & \gamma_{23}\\
0 & \gamma_{31} & \gamma_{32} & \gamma_{33}
\end{array}\right).
\]
In earlier work, considering the vicinity of a space-like singularity, the $\gamma_{ij}$ were taken to be functions of time $t$. In this work we are allowing $\gamma_{ij}$ to be functions of two parameters, time $t$ and space $x$. In both cases $\mathcal{G}$ leads to a similar structure in terms of $\gamma_{ij}$. Explicitly, for the case of two parameters,
\begin{equation}
\label{ reducedmetric}
\mathcal{G}=\frac{1}{4}\left((Tr\left[\gamma^{-1}\partial_{t}\gamma\right])^{2}+(Tr\left[\gamma^{-1}\partial_{x}\gamma\right])^{2}-Tr\left[\gamma^{-1}\partial_{t}\gamma\right]^{2}-Tr\left[\gamma^{-1}\partial_{x}\gamma\right]^{2}\right)
 \end{equation}
which reduces to the one parameter result when $\gamma_{ij}$ is a function of $t$ only.

\bibliography{stringmotiononacoset}{}
\bibliographystyle{utphys}

\end{document}